\documentclass[fleqn,usenatbib]{mnras}

\usepackage{newtxtext,newtxmath}

\usepackage[T1]{fontenc}
\usepackage{lipsum}

\DeclareRobustCommand{\VAN}[3]{#2}
\let\VANthebibliography\thebibliography
\def\thebibliography{\DeclareRobustCommand{\VAN}[3]{##3}\VANthebibliography}

\usepackage{graphicx}	
\usepackage{amsmath}	
\usepackage{orcidlink}
\usepackage{algorithm}
\usepackage{algpseudocode}
\usepackage{booktabs}
\usepackage{bbm}

\title[GPU-Accelerated Foreground Optimisation]{Optimising Foreground Modelling for Global 21cm Cosmology with GPU-Accelerated Nested Sampling}

\author[J. L. Tutt et al.]{
Jacob L. Tutt\textsuperscript{\orcidlink{0009-0002-5358-4292}},$^{1,2}$\thanks{E-mail: jlt67@cam.ac.uk} Peter H. Sims\textsuperscript{\orcidlink{0000-0002-2871-0413}},$^{1,2}$ Joe H. N. Pattison\textsuperscript{\orcidlink{0009-0006-9440-7355}},$^{1,2}$ Dominic J. Anstey\textsuperscript{\orcidlink{0000-0003-1742-7417}},$^{1,2}$ 
\and
Samuel A. K. Leeney\textsuperscript{\orcidlink{0000-0003-4366-1119}}$^{1,2}$
and 
Eloy de Lera Acedo\textsuperscript{\orcidlink{0000-0001-8530-6989}},$^{1,2}$ \\
$^{1}$Astrophysics Group, Cavendish Laboratory, J.\ J.\ Thomson Avenue, Cambridge, CB3 0HE, UK\\
$^{2}$Kavli Institute for Cosmology, Madingley Road, Cambridge, CB3 0HA, UK
}
\date{Accepted XXX. Received YYY; in original form ZZZ}

\pubyear{\the\year{}}

\begin{document}
\label{firstpage}
\pagerange{\pageref{firstpage}--\pageref{lastpage}}
\maketitle


\begin{abstract}  
The global 21-cm signal provides a powerful probe of early-Universe astrophysics, but its detection is hindered by Galactic foregrounds that are orders of magnitude brighter than the signal and distortions introduced by beam chromaticity. These challenges require accurate foreground modelling, rigorous Bayesian model comparison, and robust validation frameworks. In this work, we substantially accelerate global 21-cm inference by exploiting GPU architectures, enabling likelihood evaluations to achieve near-constant wall-clock time across a wide range of model dimensionalities and data volumes. Combined with algorithmic parallelisation of Nested Sampling, this reduces the total inference runtime of this work from hundreds of CPU-years to approximately two GPU-days, corresponding to a cost reduction of over two orders of magnitude. Leveraging this capability, we advance the physically motivated forward-modelling approach, in which foregrounds are represented by a discrete set of sky regions by introducing a novel, observation-dependent sky-partitioning scheme that defines regions using the antenna beam–convolved sky power of a given observing window. We show that this scheme improves modelling performance in three ways: firstly, by enforcing a strictly nested region hierarchy that enables clear identification of the Occam penalty in the Bayesian evidence, facilitating principled optimisation of model complexity; secondly, by enabling more accurate recovery of spatially varying spectral indices, with posterior estimates centred within physically plausible ranges; and thirdly, by allowing complex foregrounds to be modelled for robust global 21-cm signal inference using substantially fewer parameters. Overall, this approach achieves validated recovery at lower region counts, corresponding to an approximate 40\% reduction in foreground-model dimensionality.

\end{abstract}




\section{Introduction}

Driven by a wealth of high-resolution observations across the electromagnetic spectrum, the fields of astrophysics and cosmology have rapidly transformed into data-rich disciplines, allowing ever tighter constraints to be placed on the physical mechanisms and fundamental parameters governing the Universe's evolution. At high-redshift ($z \approx 1100$), observations of the Cosmic Microwave Background (CMB) \citep{COBE, WMAP, ACT, Planck2014, Planck2016, Planck2020} emitted during recombination provide a high-precision snapshot of the density anisotropies within the infant Universe. As these primordial gravitational instabilities collapsed \citep{perturbation}, they eventually resulted in the distribution of nearby galaxies within the cosmic web seen today. These low-redshift structures can be probed in similarly exquisite detail by large-scale spectroscopic surveys, such as the Baryon Oscillation Spectroscopic Survey (BOSS; \citealt{Boss}) and the Dark Energy Spectroscopic Instrument (DESI; \citealt{DESI}).

Despite these two well-explored bookends, the vast majority of cosmic history remains unmapped, most notably the Cosmic Dark Ages ($z \sim 1100-30$); the Cosmic Dawn ($z \sim 30-20$) and the Epoch of Reionisation ($z \sim 20-6$). Towards this aim, the high resolution ($\lesssim 0.1$ arcseconds) of the recently launched James Webb Space Telescope (JWST; \citealt{JWST}) is revolutionising our understanding deep within the EoR by directly imaging some of the earliest galaxies ($z \gtrsim 10$). By revealing an excess of bright ancient galaxies relative to prior predictions \citep{mismatchuv}, the mission is already prompting a re-evaluation of the theoretical models governing the Cosmic Dawn \citep[see][]{jwstsfe, JWSTIMF, jwststochasticstar}. These direct galaxy observations are complemented by a growing set of probes to provide integrated constraints on the timing, duration, and morphology of reionisation: the Ly$\alpha$ forest and Gunn--Peterson troughs seen within high-redshift quasar absorption spectra tracing the evolving opacity of the IGM \citep{Gunn_1965, Fan_2006, Becker_2015, Eilers_2018, Qin_2021, Bosman_2022}; secondary CMB anisotropies, in particular kinetic and thermal Sunyaev--Zeldovich effects sourced by the resultant ionised structures \citep{McQuinn_2005, Reichardt_2021, iliev_2024}; and line-intensity mapping surveys, which aim to measure the aggregate emission from unresolved source populations \citep{Gong_2011, Silva_2013, kovetz_2019, Ade_2020, Sun_2021, Cleary_2022, Bock_2026}.

To obtain a truly statistical picture of the IGM evolution throughout these epochs, however, a direct probe of the neutral hydrogen field itself is required. The redshifted 21-cm hyperfine transition promises to provide this, offering sensitivity to key properties such as the initial mass function \citep{gesseyjones2022}, formation efficiency, and spectral emissivity \citep{Schauer2019} of Population III stars as well as the nature of their associated X-ray binaries \citep{Sartorio_2023}  \citep[for comprehensive reviews, see][]{Furlanetto_2006, Pritchard_2012, Barkana_2016, Mesinger_2019}. Beyond these, the 21-cm signal has also been shown to be sensitive to potential exotic physics; if present, the contribution of primordial black holes \citep{Mittal_2022}, interacting dark matter models \citep{Barkana_2018}, and superconducting cosmic strings \citep{Brandenberger_2019, gesseyjones2024} would leave distinct signatures on the signal offering a unique regime to place stringent constraints on such phenomena.

Current experimental approaches for 21-cm Cosmology can be broadly classified into two categories. Firstly, interferometric arrays such as HERA \citep{HERA}, LOFAR \citep{LOFAR}, the MWA \citep{MWA} and the upcoming SKA-Low \citep{SKA} which aim to measure spatial fluctuations in the 21-cm brightness temperature through its power spectrum and, ultimately, tomographic imaging. Secondly, as a complementary approach, there exists a wealth of global 21-cm experiments focusing on the sky-averaged brightness including EDGES \citep[Experiment to Detect the Global Epoch of Reionisation Signature][]{Bowman_2018}, PRIZM \citep[Probing Radio Intensity at High-Z from Marion][]{Philip_2018}, SARAS \citep[Shaped Antenna measurement of the background RAdio Spectrum][]{Singh_2018}, MIST \citep[Mapper of the IGM Spin Temperature][]{Monsalve_2024} and REACH \citep[Radio Experiment for the Analysis of Cosmic Hydrogen][]{Acedo_2022}. While the EDGES collaboration has claimed the first tentative detection of the global 21-cm signal, its large amplitude ($\approx 500$ mK), low central frequency ($\approx 78$~MHz), and flattened Gaussian profile have been interpreted as either evidence for physics beyond standard $\Lambda$CDM cosmology \citep{Reis_2021, Liu_2019}, or as the result of residual, unmodelled systematics in the data analysis \citep{Hills_2018, Singh_2019, Sims_2019, Bevins_2021}. The latter, non-cosmological interpretation is further supported by the SARAS3's null detection, which disfavours the presence of the EDGES absorption profile in their data with 95.3 per cent confidence \citep{Singh_2022}.

\begin{figure}
    \includegraphics[width=\columnwidth, trim={0 0 0 0}, clip]{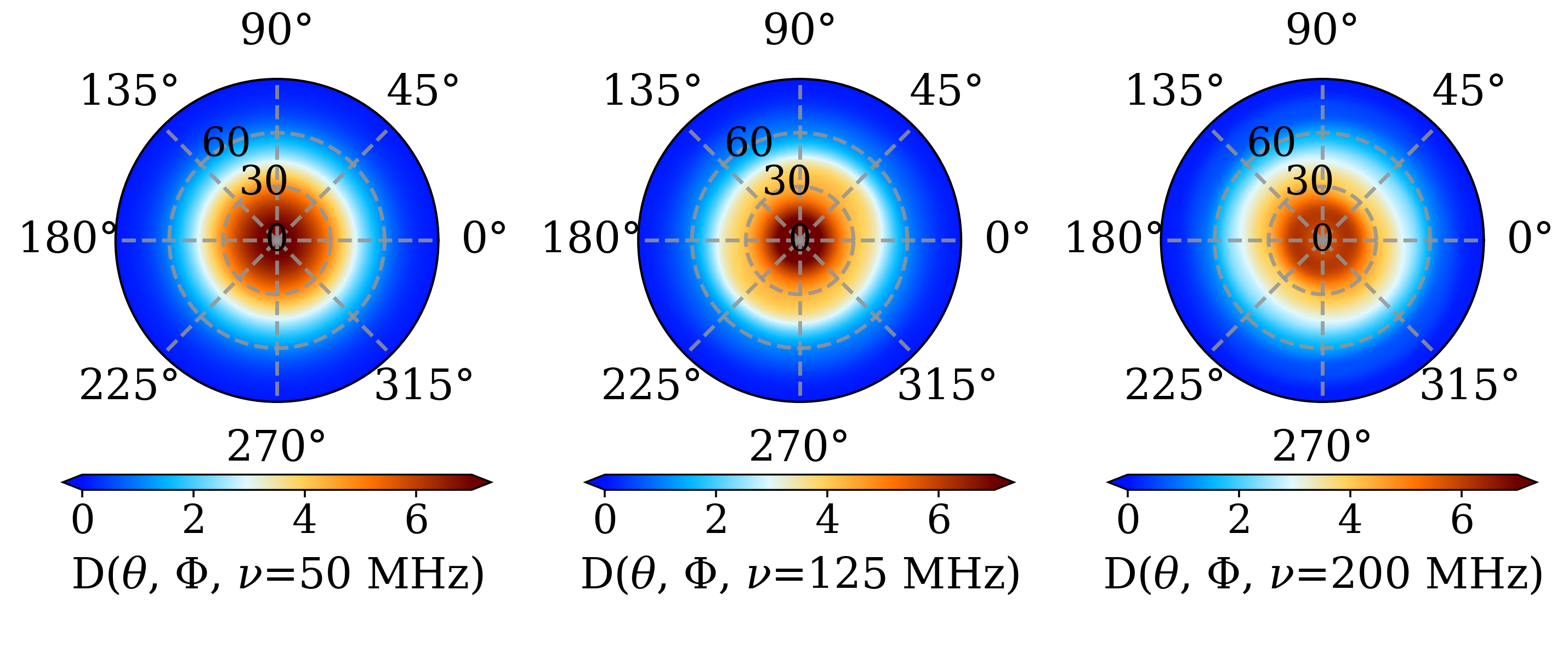}
    \caption{Beam pattern for a conical log spiral antenna, shown as polar projections of the antenna directivity $D(\theta,\phi,\nu)$ in local altitude–azimuth coordinates ($\theta,\phi$) above the horizon ($\theta < 90^\circ$). The three panels show $D(\theta,\phi,\nu)$ at $\nu$ = 50, 125 and 200~MHz to demonstrate the chromatic structure of the beam.}
    \label{fig:chromatic_beam}
\end{figure}

\begin{figure}
    \includegraphics[width=\columnwidth]{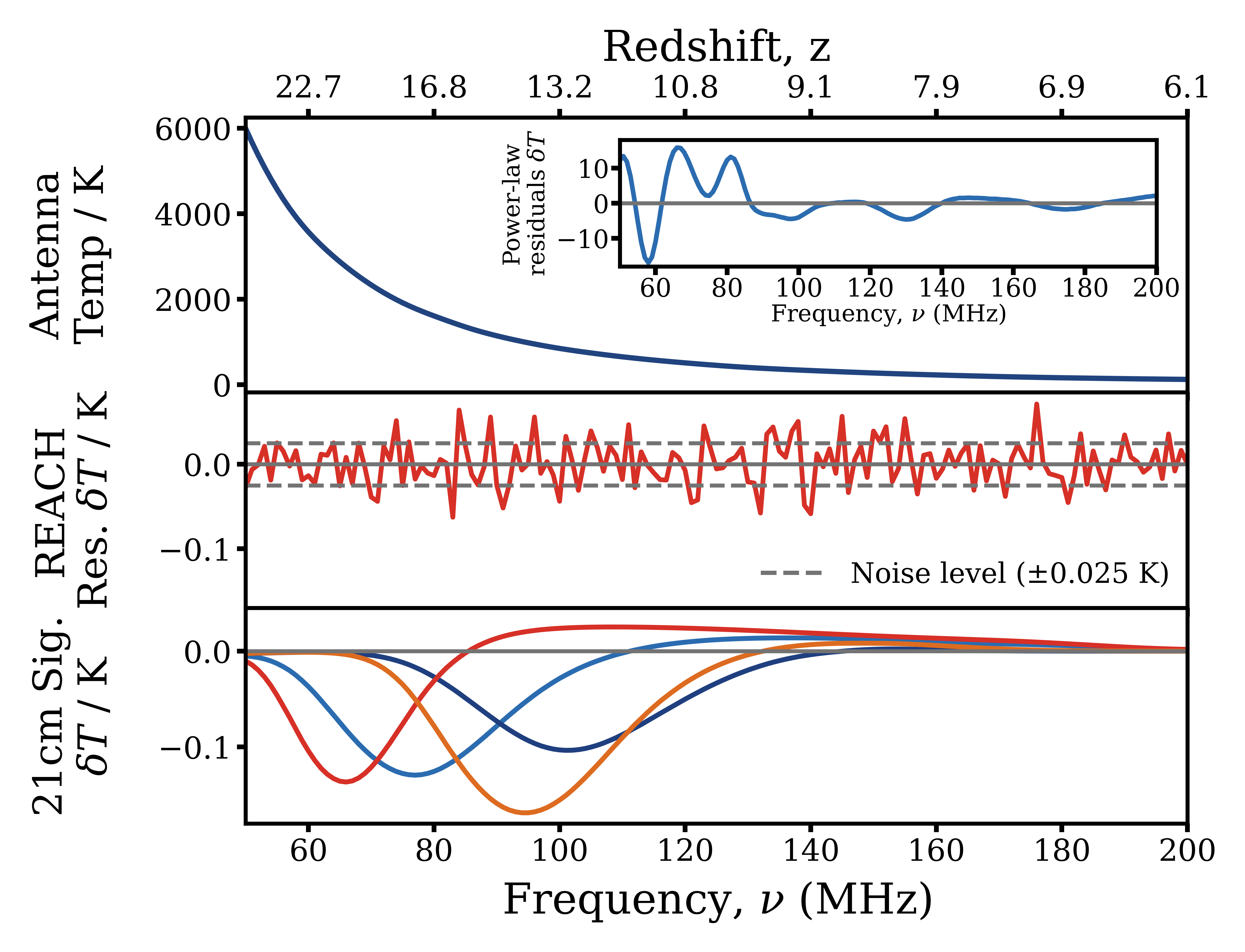}
    \caption{A demonstration of the chromatic structure introduced by the coupling of the Galactic foregrounds and the beam, its degeneracy with global 21-cm signals, and how these effects can be accounted for with the REACH data analysis pipeline. Top panel: A simulated 1 hour time-averaged observation $d(\nu)$ from the REACH telescope in the Karoo Desert, South Africa at 00:00 01-10-2019 with a conical log spiral antenna. The data has a mock 21-cm signal and 0.025~K of Gaussian noise injected. The inset shows residuals beyond a fitted smooth power-law. Middle panel: Reduced residual structure after subtracting a Bayesian nested sampling fit model produced by the REACH pipeline using 16 parametrised regions. Bottom panel: Examples of emulated mock signals from GlobalEMU \protect\citep{Bevins_2021_globalemu}, to demonstrate the success of beam-aware modelling suppressing residuals below the magnitude of expected global 21-cm signal.}
    \label{fig:chromatic_residuals}
\end{figure}

While detecting the global 21-cm signal faces a number of challenges, ranging from the mitigation of radio frequency interference (RFI; \citealt{Fridman_2001}) to distortions introduced by the ionosphere \citep{Datta_2016, Shen_2021}, one of the primary hurdles remains accurately accounting for Galactic and extragalactic foreground emission. These contaminating foregrounds exceed the expected cosmological contribution by approximately three to four orders of magnitude across the relevant frequency range ($\simeq 50$–$200$~MHz) \citep{Shaver_1999}. Typically, experiments' foreground removal strategies exploit the comparatively spectrally smooth nature of synchrotron and free–free emission, allowing them to be modelled using power laws \citep{Morales_2006}, log-polynomials \citep{Harker_2012}, or derivative-constrained functions \citep{Bevins_2021}. However, as demonstrated by \citet{Anstey_2021}, the spatial structure of Galactic foregrounds, when coupled with a chromatic antenna beam (see Figure~\ref{fig:chromatic_beam}), introduces frequency-dependent distortions that can become degenerate with the underlying cosmological signal (Figure~\ref{fig:chromatic_residuals}). In response to these challenges, the field has made significant progress across a range of complementary directions, including basis-based approaches to joint signal--foreground inference \citep{Tauscher_2018a, Rapetti_2020, Saxena_2023}, the use of foregrounds' spatial and time-dependent structure \citep{Liu_2013, Switzer_2014, Tauscher_2020a, Nhan_2019}, tests for residual structure and false-positive signal recovery \citep{Tauscher_2018b, Tauscher_2020b, Bassett_2021a, Hibbard_2023}, horizon-aware foreground modelling \citep{Bassett_2021b, Pattison_2023, pattison_2025a}, and Bayesian treatments of calibration and instrumental uncertainty \citep{Hibbard_2020, Tauscher_2021, Murray_2022}.

To enable a statistically principled inference, the REACH collaboration introduced a physically motivated, Bayesian evidence–driven analysis framework \citep{Anstey_2021, Anstey_2022} that jointly models the convolution of sky realisations with the antenna beam alongside the 21-cm signal, thereby allowing parameter degeneracies and associated uncertainties to be accurately quantified (for a demonstration see Figure~\ref{fig:chromatic_residuals}). As a consequence of this approach, the structure and spectral behaviour of the low-frequency radio sky is simultaneously constrained \citep{Carter_2025}, constituting an active area of research in its own right and a valuable resource for the broader community, independent of a confirmed detection of the global 21-cm signal \citep{deOliveiraCosta2008, Zheng_2016, Dowell_2017}.

This paper focuses on optimising parameterised sky models in a physically motivated manner in order to improve the recovery of foreground spectral index parameters and thereby better mitigate foreground systematics to levels below those that impact cosmological signal recovery. This optimisation is benchmarked using a comprehensive Bayesian validation framework \citep{Sims_2025}, enabling a rigorous comparison of model performance. Specifically, we apply this framework to the REACH radiometer \citep{Cumner_2022}, but the methodology is applicable to all physically motivated analyses of global 21-cm experiments.

The remainder of this paper is structured as follows. In Section~\ref{sec:Pipeline}, we describe the Bayesian data analysis pipeline and detail its acceleration through the parallel processing capabilities of Graphics Processing Units (GPUs) in Section~\ref{sec:gpunestedsampling}. Section~\ref{sec:regionsplittng} outlines the limitations of existing sky parameterisation approaches and introduces the observationally dependent algorithm adopted in this work. The statistical validation framework and associated performance metrics used to draw comparisons between methods are presented in Section~\ref{sec:statisticalvalidation}. Finally, the results of applying this framework to simulated datasets are reported in Section~\ref{sec:results}, and conclusions are drawn in Section~\ref{sec:conclusions}. 

\section{Bayesian Data Analysis Pipeline}
\label{sec:Pipeline}

\begin{figure*}
    \centering
    \includegraphics[width=0.85\textwidth, trim={0 0 0 2}, clip]{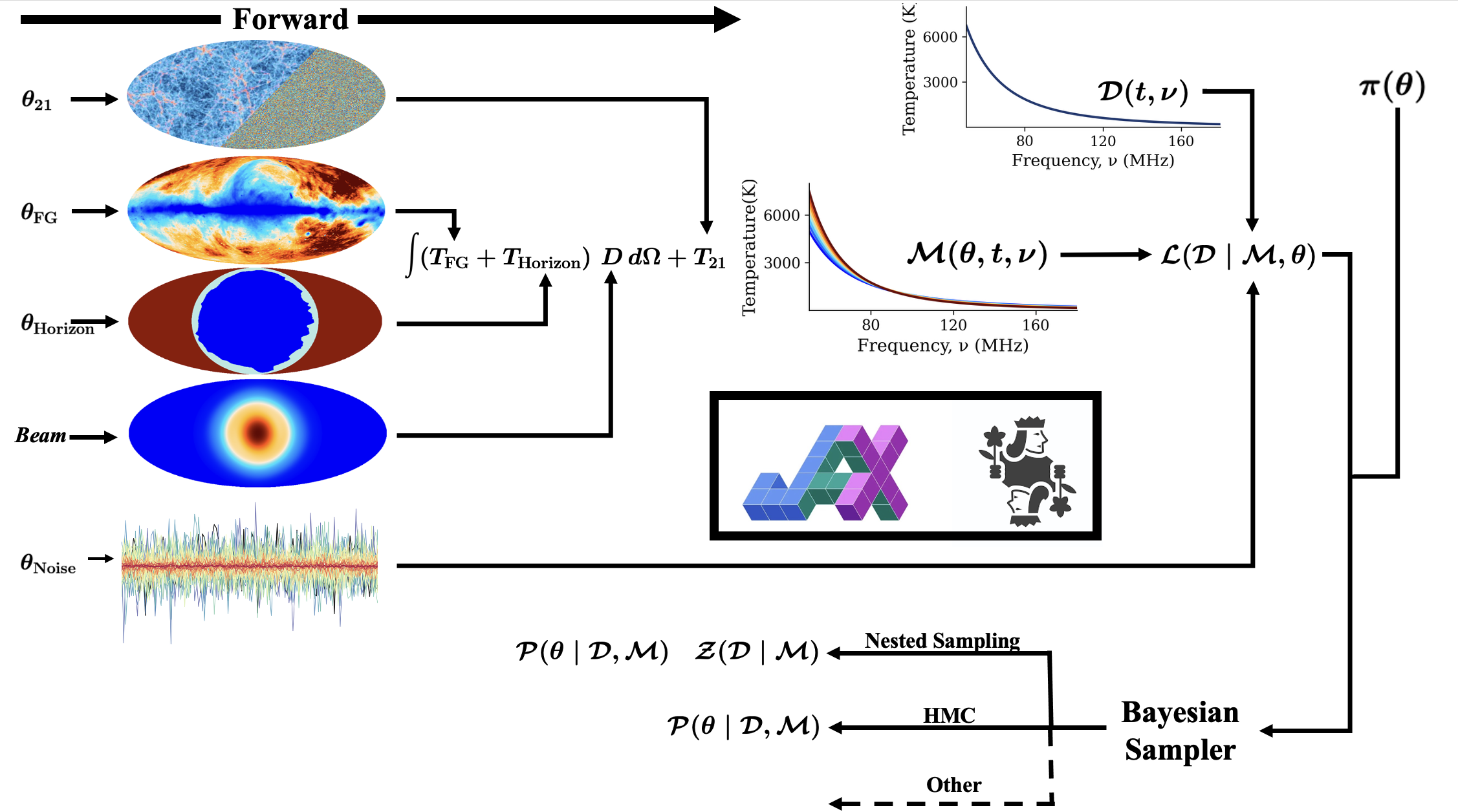}
    \caption{Schematic of the GPU-accelerated, differentiable REACH Bayesian analysis pipeline. The parameterised forward model combines a global 21-cm signal $\boldsymbol{\theta}_{\text{21cm}}$, diffuse foreground emission $\boldsymbol{\theta}_{\text{FG}}$ and horizon contamination $\theta_{\text{Horizon}}$ with the antenna's beam to generate an antenna-temperature spectrum $\mathbf{M}(\boldsymbol{\theta},t,\nu)$. This model can be statistically compared to the observational data $\mathbf{D}(t,\nu)$ under a specific noise structure through a likelihood function $\mathcal{L}(\mathbf{D}\,|\,\mathbf{M},\boldsymbol{\theta})$. The inference process is optimised through JAX’s XLA compilation \protect\cite{jax2018github}, leveraging gradient-based \texttt{BlackJAX} samplers \protect\cite{cabezas2024blackjax} and the Nested Slice Sampling (NSS) algorithm, implemented in the \texttt{BlackJAX} nested sampling framework \protect\cite{yallup2025nested}, for efficient Bayesian posterior $P(\boldsymbol{\theta} \mid \mathbf{D}, \mathbf{M})$ and evidence $\mathcal{Z}(\mathbf{D} \mid \mathbf{M})$ evaluations.}
    \label{fig:pipeline}
\end{figure*}

In this section, we present an overview of the methodology used to accurately simulate the antenna temperature for a given observation (Section~\ref{sec:DataGen}), together with the parameterised forward models (Section~\ref{sec:ForwardModel}) and Bayesian inference sampling algorithms (Section~\ref{sec:bayesianinference}) employed to efficiently solve the associated inverse problem. While the discussion below focuses on accounting for the diffuse foregrounds alongside the cosmological signal, the mathematical formalism required to incorporate additional physical effects, including RFI \citep{Leeney_2023, Anstey_2024}, the ionosphere \citep{Shen_2022}, extragalactic point sources \citep{Mittal_2024} and environmental conditions \citep{Pattison_2023, pattison_2025a} is already established and can be incorporated in a modular manner. A schematic overview of the full analysis pipeline is shown in Figure~\ref{fig:pipeline}.

\subsection{Data Simulation}
\label{sec:DataGen}

To fully capture the chromatic distortions introduced by diffuse foreground emissions into the observed data, simulations require full-resolution sky models that encode realistic spatial structure and frequency-dependent intensity distributions. These models will hereafter be referenced in relation to the local Alt–Az ($\theta,\phi$) coordinate frame of the antenna and thus a function of Coordinated Universal Time (UTC). Following \citealt{Anstey_2021}, an observationally motivated realisation of such a model can be obtained using the 2008 Global Sky Model \citep[GSM][]{deOliveiraCosta2008} evaluated at 408~MHz ($T_{408}(\theta,\phi, t)$) and 230~MHz ($T_{230}(\theta,\phi, t)$). By comparing the two frequencies, a spatially varying spectral index field $\beta(\theta,\phi,t)$ is derived as:

\begin{equation} 
\label{eqn:spectralindexmap}
\beta(\theta, \phi, t)=
\frac{
\log\!\left[(T_{230}(\theta,\phi, t) - T_{\mathrm{CMB}}) / (T_{408}(\theta,\phi, t) - T_{\mathrm{CMB}})\right]
}{
\log\!\left(230/408\right)
},
\end{equation}

\noindent This spectral-index field is then used to extrapolate a reference sky at frequency $\nu_0$ (taken here to be $T_{230}(\theta,\phi, t)$) to arbitrary observing frequencies:

\begin{equation}
\label{eqn:skyapproximation}
T_{\mathrm{sky}}(\theta, \phi, \nu, t)
=
\left[T_{230}(\theta,\phi,t) - T_{\mathrm{CMB}}\right]
\left(\frac{\nu}{\nu_0}\right)^{-\beta(\theta,\phi,t)}
+ T_{\mathrm{CMB}}.
\end{equation}

\noindent This procedure yields a continuous low-frequency sky model $T_{\mathrm{sky}}(\theta,\phi,\nu,t)$, which, after convolution with the instrumental response, defines a continuous antenna-temperature field. In practice, this field is evaluated and stored over the finite frequency channels and Local Sidereal Time (LST) bins of the observation. The data, forward model and likelihood can therefore be represented on the resulting frequency--time grid as matrices.

\noindent The simulated sky brightness temperature is then convolved with the directivity of the antenna beam, $D(\theta,\phi,\nu)$, to produce the corresponding antenna temperature $T_{\mathrm{data}}(\nu,t)$. For demonstration purposes through this work, the antenna beam is modeled as a 6-m conical log-spiral antenna (Figure~\ref{fig:chromatic_beam}). To facilitate signal recovery tests, the mock data includes a realistic mock global 21-cm signal, $T_{21}(\nu)$. This is modeled as a Gaussian absorption profile, $\mathbf{M}_{21}(\theta_{21})$, with a central frequency ($\nu_{21}$) of 85~MHz, a 15~MHz width ($\sigma_{21}$), and an amplitude of 0.155~K ($A_{21}$):

\begin{equation}
\label{eqn:t21model}
T_{21}(\nu)
=
A_{21}\,
\exp\!\left[
-\frac{(\nu - \nu_{21})^{2}}{2\sigma_{21}^{2}}
\right].
\end{equation}

\noindent Finally, an assumed noise realisation $\hat{\sigma}$ is added, which for the purposes of being comparable with prior work was chosen to be uncorrelated Gaussian noise with an amplitude of 25~mK:

\begin{equation}
\label{eqn:simantenna}
T_{\mathrm{data}}(\nu, t)
=
\frac{1}{4\pi}
\int_{0}^{4\pi}
D(\theta,\phi,\nu)\,
T_{\mathrm{sky}}(\theta,\phi,\nu,t)\,
\mathrm{d}\Omega
+ T_{21}(\nu)
+ \hat{\sigma}.
\end{equation}

\noindent Throughout this work, we benchmark our parameterised models against a range of simulated foreground complexities derived from three distinct observational windows with varying Galactic orientations, as detailed in \autoref{tab:obs_windows}. These shorter integration periods represent challenging modelling scenarios where the chromatic structure induced by the lack of sky rotation overhead is significant. This is particularly pronounced in the `Galaxy Up' case, where the power of the Galactic centre further magnifies the amplitude of the distortions introduced. Consequently, the `Galaxy Up' case represents an extreme scenario where 21-cm signal recovery would likely not be attempted in isolation on real data. However, it serves as a robust stress test of our dynamic model's ability to improve foreground recovery, and ensures that our validation metrics correctly flag reconstructions that are inadequate for the precision required for cosmological inference. Conversely, the 4-hour integration spans a broader range of Galactic positions, representing a more typical and viable observation target for signal inference. We note that while the equations presented throughout this section maintain the time-dependent $t$ notation for mathematical generality, as time-resolved analysis is benchmarked in Section~\ref{app:TimingGPU}, the primary analysis in this work is performed on time-integrated data. This is achieved by averaging over the time domain in both the forward models and data simulations, creating a single integrated spectrum for each observational window.

\begin{table}
\centering
\caption{Parameters for the three observational windows used to benchmark foreground reconstruction and signal recovery, all starting at 00:00:00 on the respective date.}
\label{tab:obs_windows}
\begin{tabular}{l l l}
\hline \hline
\textbf{Reference} & \textbf{Date/ Duration} & \textbf{Configuration} \\ \hline
Galaxy Down & 01-10-2019 / 1 hr & Galactic centre below horizon. \\
Galaxy Up   & 01-07-2019 / 1 hr & Galactic centre at zenith. \\
Galaxy 4hr  & 01-01-2019 / 4 hr & Integrated transit of the Galaxy. \\ \hline
\end{tabular}
\end{table}

\subsection{Physically Motivated Foreground Model}
\label{sec:ForwardModel}

While the simulation pipeline described in Section~\ref{sec:DataGen} provides high-fidelity realisations, the calculated spatial distribution of both the base temperature $T_{408}(\theta,\phi,t)$ and the spectral index $\beta(\theta,\phi,t)$ is subject to observational uncertainties and thus likely offset from the true radio foregrounds. When aiming to perform signal inference from observational data, using these as fixed templates would introduce systematics into the modelled antenna temperature that prevent the recovery of the true global 21-cm signal. To mitigate this, the spectral indices and base-map amplitude must be simultaneously parameterised and incorporated within the foreground model to be jointly fit (see similiar interferometric treatments in \citet{Sims_2016,sims_2019a,sims_2019b,Sims_2022a,Sims_2022b,burba_2023,Wilensky_2025,kern_2025}).

However, due to the nature of the one-dimensional data $T_\mathrm{data}(\nu)$ or two-dimensional data $T_\mathrm{data}(\nu, t)$ produced by a single radiometer, full pixel-level parameterisation would be highly degenerate and computationally prohibitive. Therefore, we adopt a regional parameter approach \citep{Anstey_2021,Anstey_2022,Pagano_2023}, which partitions the sky into $N_{\beta}$ regions of uniform spectral index and $N_{\alpha}$ regions of uniform amplitude scaling. These are defined by the independent binary masks, $M_{\beta,j}$ and $M_{\alpha,i}$, which determine the fixed membership of a pixel ($\theta, \phi, t$) to a specific region:

\begin{equation}
\label{eqn:forward_model}
\begin{split}
T_{\mathrm{sky}}^{\mathrm{model }} (\theta, \phi, \nu, t& ) = \sum_{i=1}^{N_{\alpha}} \sum_{j=1}^{N_\beta}  M_{\alpha,i}(\theta, \phi, t) M_{\beta,j}(\theta, \phi, t) \\
& \times \left[ \alpha_i \left( T_{408}(\theta, \phi, t) - T_{\mathrm{CMB}} \right) \right] \left( \frac{\nu}{\nu_0} \right)^{-\beta_j} + T_{\mathrm{CMB}},
\end{split}
\end{equation}

\noindent where $\alpha_i$ is a multiplicative scale factor effectively acting as a localised gain correction to the 230~MHz template, and $\beta_j$ is the fit spectral index for the $j^{\mathrm{th}}$ region.

To accelerate the forward model for a given observation window $\boldsymbol{t}$ and frequency band $\boldsymbol{\nu}$, the interaction between the beam, regional masks, and the reference base-map during the sky integration can be precomputed. This defines a set of chromatic response functions, $\mathcal{K}_{i,j}(\nu, t)$, which allows the resultant antenna temperature, conditioned on any set of foreground parameters $\{\boldsymbol{\alpha}, \boldsymbol{\beta}\}$, to be reduced to a simple series of matrix operations,

\begin{equation} 
\label{eqn:chromatic_function}
\begin{split}
\mathcal{K}_{i,j}(\nu, t) = \frac{1}{4\pi} \int^{4\pi} & M_{\alpha,i}(\theta, \phi, t) M_{\beta,j} (\theta, \phi, t)  \\ & \times \left[ T_{408}(\theta, \phi, t) - T_{\mathrm{CMB}} \right] D(\theta, \phi, \nu)  \mathrm{d}\Omega\, ,
\end{split}
\end{equation}

\noindent and hence the antenna temperature is given by:

\begin{equation} 
\label{eqn:fast_forward_model} 
T_{\mathrm{FG}}(\nu, t) = \sum_{i=1}^{N_{\alpha}} \sum_{j=1}^{N_\beta} \alpha_i \,  \mathcal{K}_{i,j}(\nu, t) \left( \frac{\nu}{\nu_0} \right)^{-\beta_j} + T_{\mathrm{CMB}}.
\end{equation}

\noindent While previous implementations of this approach have relied on relatively simple partitioning schemes, this work focuses on optimising region definitions (Section~\ref{sec:regionsplittng}) to reconstruct continuous foregrounds with a minimal parameter set. This refinement is constrained by the condition that the models remain sufficiently expressive to ensure any residual systematics are statistically insignificant relative to the noise structure, a condition verified through a comprehensive validation suite (Section~\ref{sec:statisticalvalidation}). In the interest of clarity, this paper focuses specifically on the definition of spectral index masks ($M_{\beta,j}$); however, the algorithms introduced in Section~\ref{sec:regionsplittng} are equally applicable to the amplitude scale factor masks ($M_{\alpha,i}$), the demonstration of which is left for future work.

\subsection{Bayesian Inference}
\label{sec:bayesianinference}

Given a forward model $\mathbf{M}$ describing the foregrounds and redshifted 21-cm signal through a set of parameters $\boldsymbol{\theta}_{\mathbf{M}}$, we employ Bayesian inference to perform parameter estimation and model comparison for a given observational dataset $\mathbf{D}$. This is achieved by applying Bayes’ theorem:

\begin{equation}
\label{eqn:bayes_theorem}
P(\boldsymbol{\theta}_{\mathbf{M}} \mid \mathbf{D}, \mathbf{M}) = \frac{\mathcal{L}(\mathbf{D} \mid \boldsymbol{\theta}_{\mathbf{M}}, \mathbf{M}) \, \, \pi(\boldsymbol{\theta}_{\mathbf{M}} \mid \mathbf{M})}{\mathcal{Z}(\mathbf{D} \mid \mathbf{M})},
\end{equation}

\noindent where $\pi(\boldsymbol{\theta}_{\mathbf{M}} \mid \mathbf{M})$ denotes the prior probability distribution over the model parameters, encoding our initial state of knowledge (or lack thereof). The likelihood, $\mathcal{L}(\mathbf{D} \mid \boldsymbol{\theta}_{\mathbf{M}}, \mathbf{M})$, quantifies the probability of obtaining the observed data given a specific forward model, parameter set, and assumed noise structure. The resulting posterior, $P(\boldsymbol{\theta}_{\mathbf{M}} \mid \mathbf{D}, \mathbf{M})$, represents the updated probability distribution of parameters after incorporating the information contained in the data. Finally, $\mathcal{Z}(\mathbf{D} \mid \mathbf{M})$ is the Bayesian evidence, which measures the overall support for a model given the data and is defined as the likelihood marginalised over the full prior volume of parameters:

\begin{equation}
\label{eqn:evidence}
\mathcal{Z}(\mathbf{D} \mid \mathbf{M}) = \int \mathcal{L}(\mathbf{D} \mid \boldsymbol{\theta}_{\mathbf{M}}, \mathbf{M}) \, \pi(\boldsymbol{\theta}_{\mathbf{M}} \mid \mathbf{M}) \, \mathrm{d} \boldsymbol{\theta}_{\mathbf{M}}.
\end{equation}

\subsubsection{Likelihood Function}

In this work, we assume the noise $\hat{\sigma}$ is adequately described by a homoskedastic Gaussian distribution, however a host of more complex noise structures including radiometric noise \citep{ Scheutwinkel_2022a, Scheutwinkel_2022b} have been previously explored in the context of 21-cm signal recovery. Given a dataset sampled across frequencies $\boldsymbol{\nu}$ and times $\boldsymbol{t}$, the log-likelihood ($\ln \mathcal{L}$) can thus be expressed as:

\begin{equation}
\label{eqn:likelihood}
\ln \mathcal{L} = -\frac{1}{2} \sum_{k,l} \left[ \ln(2\pi\sigma_n^2) + \frac{\left( \mathbf{T}_{\mathrm{data}}(\nu_k, t_l) - \mathbf{M}(\nu_k, t_l, \boldsymbol{\theta}) \right)^2}{\sigma_n^2} \right]
\end{equation}

\noindent where $\mathbf{M}(\nu_k, t_l, \theta) = \mathbf{T}_{\mathrm{FG}}(\nu_k, t_l,  \boldsymbol{\theta_{\mathrm{FG}}}) + \mathbf{T}_{21}(\nu_k, \boldsymbol{\theta_{21}})$ represents the combined foreground and signal forward model. Additionally, we fit the Gaussian noise standard deviation, $\sigma_n$, as a free parameter during inference.

\subsubsection{Prior Distributions}

The prior distributions utilised throughout our primary analysis (\autoref{tab:priors}) were chosen to be sufficiently broad to encompass physically plausible realisations of the low-frequency sky. The spectral index priors were bounded by the extremes of the $\beta(\theta, \phi, t)$-map derived in Equation~\ref{eqn:spectralindexmap}. For the global 21-cm signal, the prior ranges are informed by non-exotic astrophysical simulations \citep{Cohen_2017} from the semi-numerical code 21cmSPACE \citep{Fialkov_2013, Fialkov_2014}, reflecting the expected structure of signals.

\begin{table}
\centering
\caption{Prior distributions for the global 21-cm signal, regional foregrounds, and instrumental noise parameters.}
\label{tab:priors}
\begin{tabular}{l c c c}
\hline \hline
\textbf{Parameter} & \textbf{Prior Dist.} & \textbf{Range} & \textbf{Units} \\ \hline
\textit{Statistical Noise} & & & \\
Noise Amplitude ($\sigma_n$) & Log-Uniform & $[10^{-4}, 10^{-1}]$ & K \\ \hline
\textit{Regional Foregrounds} & & & \\
Spectral Index ($\beta_j$) & Uniform & $[2.458, 3.146]$ & -- \\ \hline
\textit{Global 21-cm Signal} & & & \\
Amplitude ($A_{21}$) & Uniform & $[0, 0.25]$ & K \\
Centre Frequency ($\nu_{21}$) & Uniform & $[50, 200]$ & MHz \\
Width ($\sigma_{21}$) & Uniform & $[10, 20]$ & MHz \\ \hline
\end{tabular}
\end{table}

\subsubsection{Model Selection}
\label{sec:modelselection}

Given the extreme sensitivity of global 21-cm signal recovery to mismodelling, it is critical to quantitatively demonstrate that any observational dataset $\mathbf{D}$ used for a claimed detection most strongly supports a signal-plus-foreground model ($\mathbf{M}_i$) against a foreground-only or alternative residual systematics model ($\mathbf{M}_j$). In order to perform this robust model comparison, evaluating the Bayesian evidence is essential as the relative probability of two competing models, i.e., $\mathbf{M}_i$ and $\mathbf{M}_j$, is determined by the ratio of their posterior odds, $R_{ij}$. By applying Bayes' theorem at the model level and assuming a non-informative (uniform) prior belief between them, $\pi(\mathbf{M}_i) = \pi(\mathbf{M}_j)$, this reduces to the ratio of their evidences, known as the Bayes Factor, $B_{ij}$:

\begin{equation}
\label{eqn:bayes_factor}
R_{ij} = \frac{P(\mathbf{M}_i \mid \mathbf{D})}{P(\mathbf{M}_j \mid \mathbf{D})} 
= \underbrace{\frac{\mathcal{Z}(\mathbf{D} \mid \mathbf{M}_i)}{\mathcal{Z}(\mathbf{D} \mid \mathbf{M}_j)}}_{B_{ij}} \, \, \underbrace{\frac{\pi(\mathbf{M}_i)}{\pi(\mathbf{M}_j)}}_{\text{prior odds}}.
\end{equation}

\noindent To interpret the quantitative strength of preference implied by the Bayes factor $B_{ij}$ (or equivalently $R_{ij}$), we adopt the qualitative classification scheme established by \citet{Kass_1995}, under which the degree of support for a given comparative model is categorised according to the ranges summarised in \autoref{tab:jeffreys}.

While gradient-based Markov Chain Monte Carlo (MCMC) methods, such as Hamiltonian Monte Carlo, excel at efficiently exploring the posterior topology to identify parameter correlations \citep{Duane_1987, Neal_1996, Hoffman_2014}, they do not natively provide an efficient means to calculate the Bayesian evidence, $\mathcal{Z}$. To address this, in this work we employ Nested Sampling \citep[NS,][]{Skilling_2006}, which simultaneously yields posterior samples and an accurate estimate of the evidence, as outlined in Section~\ref{sec:gpunestedsampling}.

\begin{table}
\centering
\caption{Interpretive mapping between the log Bayes factor $\ln(B_{ij})$ and qualitative levels of support for model $\mathbf{M}_i$ relative to $\mathbf{M}_j$.}
\label{tab:jeffreys}
\begin{tabular}{lll}
\hline \hline
\textbf{ln($B_{ij}$)} & \textbf{Odds in favour of $\mathbf{M}_i$} & \textbf{Preference for $\mathbf{M}_i$} \\ \hline
$0 \leq \ln(B_{ij}) < 1$ & $1-3$     & Weak      \\
$1 \leq \ln(B_{ij}) < 3$ & $3-20$    & Moderate  \\
$3 \leq \ln(B_{ij}) < 5$ & $20-150$  & Strong    \\
$\ln(B_{ij}) \geq 5$     & $> 150$   & Decisive  \\ \hline
\end{tabular}
\end{table}

\section{GPU-Accelerated Nested Sampling}
\label{sec:gpunestedsampling}

While Nested Sampling's approach to evaluating the marginal likelihood (Equation~\ref{eqn:evidence}) is widely adopted across astrophysics and cosmology, as the dimensionality of the parameter space and the volume of observational data increases traditional CPU-based implementations face significant scalability challenges. In this work alone, the systematic exploration and validation of various foreground partitioning schemes (see Section~\ref{sec:statisticalvalidation}) necessitates $\mathcal{O}(10^3)$ independent nested sampling runs, corresponding to a total computational cost of order $\sim 10^6$ CPU-hours (see Section~\ref{app:Likelihoodaccel} for hardware configuration details). Given this, reducing the overall inference cost in terms of wall-time and computational resources is essential for feasible and reproducible analysis. 

To address this, we integrate the Nested Slice Sampling (NSS) algorithm, implemented in the \texttt{BlackJAX} nested sampling framework, into the analysis pipeline \citep{yallup2025nested, cabezas2024blackjax}. This enables the exploitation of the parallel architecture of Graphics Processing Units (GPUs) at two complementary levels: firstly, hardware-level vectorisation of likelihood evaluations, detailed in Section~\ref{sec:Likeparrallel}; secondly, algorithmic reformulation of the nested sampling procedure, outlined in Section~\ref{sec:algoparrallel}. The discussion below summarises the NSS methodology introduced by \citet{yallup2025nested}, while benchmarking of the resulting performance gains is presented in Section~\ref{app:TimingGPU}. Although not a primary focus of this work, implementing the analysis pipeline within a \texttt{JAX}- and GPU-compatible framework also enables automatic differentiation \citep{autodiff_2017}, allowing gradients of the forward model and likelihood to be obtained at negligible additional computational cost. This naturally opens the door to future extensions within the nested sampling framework, including gradient-informed nested sampling schemes \citep{Betancourt_2010, Lemos_2024} and Slice-within-Gibbs nested sampling kernels for hierarchical models \citep{yallup2026nestedsamplingslicewithingibbsefficient}. Beyond nested sampling, other scalable evidence-estimation strategies include Sequential Monte Carlo methods with Hamiltonian Monte Carlo inner kernels \citep{buchholz2020adaptivetuninghamiltonianmonte}, as well as the integration of physically motivated forward models within Bayesian machine learning pipelines \citep[see][and Leeney et al., in prep.]{saxena_2024}.

\subsection{Acceleration Mechanism}

\subsubsection{Likelihood Parallelisation}
\label{sec:Likeparrallel}

As outlined in Section~\ref{sec:ForwardModel}, following appropriate pre-computation, such as the construction of chromatic response functions $\mathcal{K}_{i,j}(\nu,t)$, the likelihood reduces to a sequence of batched linear algebra operations. This structure maps naturally onto GPUs, which are explicitly designed to perform large-scale linear algebra workloads efficiently.

GPUs comprise thousands of lightweight cores optimised for high-throughput, Single Instruction–Multiple Data (SIMD) workloads, in contrast to CPUs, which prioritise low-latency, sequential performance \citep{Owens_2008}. When coupled with \texttt{JAX}'s Accelerated Linear Algebra (XLA) compilation \citep{Sabne_2020}, they allow likelihood evaluations to be executed concurrently across many threads, yielding substantial reductions in wall-time. As a result, the effective scaling with increasing data volume or model dimensionality is strongly suppressed, approaching $\mathcal{O}(1)$ behaviour, until limited by memory bandwidth.

\subsubsection{Algorithmic Parallelisation}
\label{sec:algoparrallel}

The second level of acceleration involves an adaptation of the Nested Sampling algorithm itself. Originally proposed by \citet{Skilling_2006}, NS solves the evidence integral by mapping the high-dimensional parameter space $ \boldsymbol{\theta}$ to a one-dimensional prior mass $X$, defined as the fractional volume of the prior contained within an iso-likelihood contour, $\mathcal{L}^*$:
\begin{equation}
X(\mathcal{L}^*) = \int_{\mathcal{L}(\boldsymbol{\theta}) > \mathcal{L}^*} \pi(\boldsymbol{\theta})\,\mathrm{d}\boldsymbol{\theta}.
\end{equation}

\noindent Under this transformation, the evidence is simply reduced to a one-dimensional integral over prior mass. In conventional CPU implementations, this integral is evaluated numerically by evolving a population of $n_{\mathrm{CPU}}$ live points, with each iteration contracting the remaining prior mass through replacement of the point with the lowest likelihood. This process generates a discrete set of discarded (dead) points which allow the integral to be approximated via a weighted summation:
\begin{equation}
\label{eqn:evidence_sum}
\mathcal{Z} = \int_{0}^{1} \mathcal{L}(X)\,\mathrm{d}X \approx \sum_{i} (X_{i-1} - X_{i})\mathcal{L}_{i}.
\end{equation}
Here, $\mathcal{L}_{i}$ is the likelihood of the $i^{\text{th}}$ discarded point, and the quadrature weight $(X_{i-1} - X_{i})$ is determined by the stochastic contraction of the prior mass. For a constant population of $n_{\mathrm{CPU}}$ live points, the expected log-volume remaining after $k$ iterations is given by:
\begin{equation}
\label{eqn:log_xcpu}
\mathbb{E}\left[\log X_{\mathrm{CPU}}\right] = -\sum_{j=1}^{k} \frac{1}{n_{\mathrm{CPU}}} = -\frac{k}{n_{\mathrm{CPU}}}.
\end{equation}

\noindent Following \citet{yallup2025nested}, this inherently sequential process can be parallelised by simultaneously discarding the $b$ lowest-likelihood points and launching independent Markov chains in parallel across the GPU, each subject to the constraint $\mathcal{L} > \mathcal{L}_{\text{min}, b}$, where $\mathcal{L}_{\text{min}, b}$ is the maximum likelihood of the discarded batch. Accounting for the effective decrease in the number of live points throughout the batch, the expectation of the cumulative log-volume contraction of one parallelised run is then:

\begin{equation}
\label{eqn:log_xgpu}
\mathbb{E}\left[ \log X_{\text{GPU}}\right] = -\sum_{i=0}^{b-1} \frac{1}{n_{\text{GPU}} - i} \approx \ln \left( \frac{n_{\text{GPU}} - b}{n_{\text{GPU}}} \right).
\end{equation}

\noindent For further demonstrations of GPU-accelerated nested sampling applied to a broader range of cosmological and astrophysical inference problems, we refer the reader to \citet{yallup_2025_gw, prathaban_2025, leeney_2025_bandflux, leeney_2025_anomoly} and \citet{lovick_2025}. 

\subsubsection{Sampler Hyperparameters}
\label{sec:hyperparam}

The performance of the sampling algorithm is governed by a small number of key hyperparameters. To generate new, approximately uncorrelated live points, we use Nested Slice Sampling (NSS) as implemented in \texttt{BlackJAX} \citep{yallup2025nested}. In this scheme, each replacement begins from a randomly selected live point and evolves a hit-and-run slice-sampling chain within the likelihood-constrained region, rather than by slice sampling in the unit hypercube. At each slice step, the proposal direction is drawn in a Mahalanobis-normalised geometry defined by a covariance-based whitening transform fitted to the existing live-point population, which improves exploration of correlated parameter spaces and supports a near-unity step size in the whitened coordinates.

The efficiency and accuracy of this process are primarily controlled by the number of live points used throughout the run (\texttt{n\_live}), which influences the resolution of the posterior and evidence estimation; the number of slice-sampling steps used to generate new live points (\texttt{num\_inner\_steps}), which regulates the degree of correlation between samples; and the number of live points replaced simultaneously during each iteration (\texttt{num\_delete}), which determines the level of algorithmic parallelism. This final parameter effectively trades computational efficiency against sampling accuracy. Throughout this work, we adopt \texttt{n\_live} = $400 \times N_{\mathrm{Dim}}$, \texttt{num\_inner\_steps} = $12 \times N_{\mathrm{Dim}}$, and \texttt{num\_delete} = $0.2 \times \texttt{n\_live}$, where $N_{\mathrm{Dim}}$ denotes the number of free parameters in the given model configuration. These settings are motivated by the convergence and performance studies of Pattison et al. (in prep.).

\subsection{Performance Benchmarking}
\label{app:TimingGPU}

\begin{figure} 
    \centering
    \includegraphics[width=\columnwidth, trim=0 0 0 0, clip]{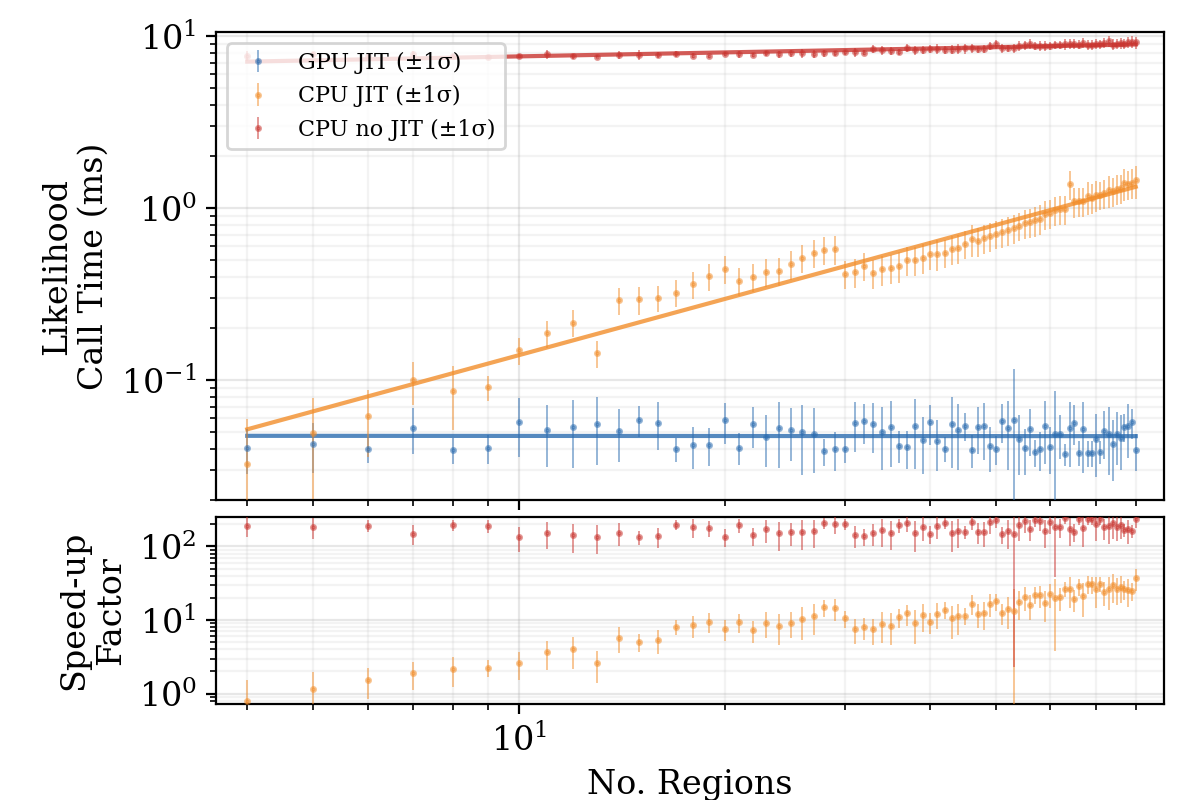}
    \caption{Performance benchmarking of the likelihood evaluation across varying model complexities. Top: Comparison of mean execution time (ms) for 1000 likelihood calls on an Intel Cascade Lake CPU (with and without JIT compilation) versus an NVIDIA A100 GPU. Bottom: The resulting speed-up factor of the A100 implementation relative to both JIT and non-JIT CPU baselines.}
    \label{fig:likelihood_timing_dim}
\end{figure}

Having established the two complementary acceleration pathways, we now benchmark the performance gains associated with each stage by comparing GPU-accelerated implementations against traditional CPU-based approaches. We focus on reductions in wall-clock time and overall financial cost, first at the likelihood-evaluation level and then for end-to-end nested sampling runs.

\subsubsection{Likelihood Acceleration}
\label{app:Likelihoodaccel}

We first evaluate the execution time of the likelihood function on both CPU and GPU architectures, examining how performance scales with the dimensionality of the parameter space (controlled by the number of regions) and with data volume. When investigating the former, to isolate the independent contributions of speed-up from compiler optimisation and hardware parallelism, we initially benchmark the effect of just-in-time (JIT) compilation on CPU execution, before showing the maximal effect on an NVIDIA A100 GPU. The two distinct stages of computational efficiency are evident from the resulting performance trends shown in Figure~\ref{fig:likelihood_timing_dim}.

JIT compilation alone yields an improvement of approximately two orders of magnitude in the constant computational overhead of the likelihood evaluation. Furthermore, while both the compiled and uncompiled CPU implementations scale approximately linearly with model dimensionality, $\mathcal{O}(N)$, JIT compilation significantly reduces the magnitude of this scaling, from an increase of $0.023\,\mathrm{ms}$ per additional region in the uncompiled case to $0.017\,\mathrm{ms}$ per region for the compiled. This demonstrates that compiler optimisation substantially accelerates the sequential CPU execution, both by reducing fixed overheads and by improving scaling behaviour.

\begin{figure} 
    \centering
    \includegraphics[width=\columnwidth, trim=0 0 0 0, clip]{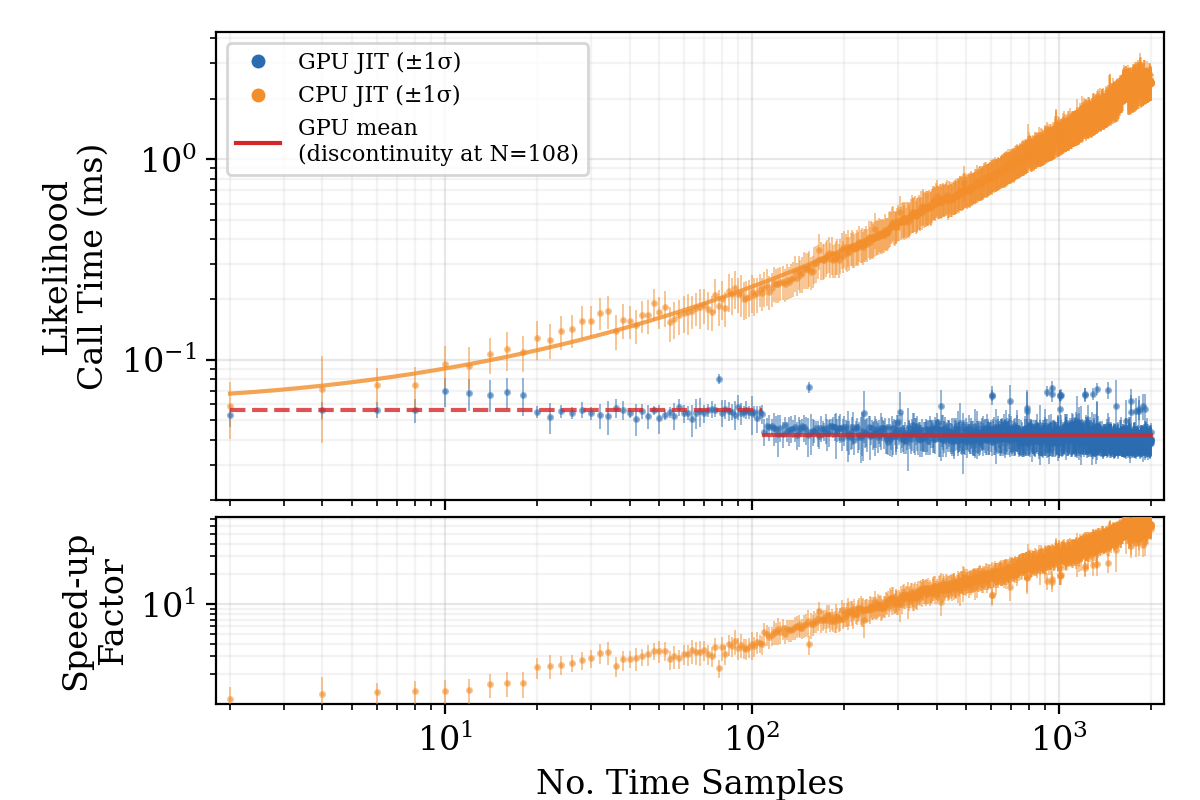}
    \caption{Performance benchmarking of the likelihood evaluation across varying data volumes. Top: Comparison of mean execution time (ms) for 1000 likelihood calls on an Intel Cascade Lake CPU (with JIT compilation) versus an NVIDIA A100 GPU, with discontinuity highlighted in by red line (dashed/solid). Bottom: The resulting speed-up factor of the A100 implementation relative to the JIT CPU baselines.}
    \label{fig:likelihood_timing_data}
\end{figure}

\begin{table*}
\centering
\small
\begin{tabular}{lccccc}
\toprule
Configuration & No. Regions & GPU Runtime (s) & CPU Runtime (s) & Speed-up Factor & Price Factor \\
\midrule
Integrated & 10 & 25.48 & 130.36 & 5.12 & 2.05 \\
Integrated & 20 & 39.68 & 568.76 & 14.33 & 5.73 \\
Integrated & 30 & 72.71 & 2245.52 & 30.88 & 12.35 \\
\midrule
Resolved   & 10 & 39.04 & 409.68 & 10.49 & 4.20 \\
Resolved   & 20 & 123.19 & 3752.14 & 30.46 & 12.18 \\
Resolved   & 30 & 315.13 & 16611.17 & 52.71 & 21.08 \\
\bottomrule
\end{tabular}
\caption{Comparison of runtime and cost efficiency between \texttt{BlackJAX} executed on an NVIDIA A100 GPU and \texttt{PolyChord} executed on a 40-core Intel Ice Lake CPU. Runtimes are evaluated for a fixed six-hour observation window corresponding to 72 observations. The speed-up factor is defined as $t_{\mathrm{CPU}}/t_{\mathrm{GPU}}$, while the price factor denotes the relative financial cost per run based on hardware pricing on the Cambridge Service for Data Driven Discovery (CSD3) HPC system, with costs of 40\,p per hour for the 40-core CPU node and 100\,p per A100 GPU-hour.}
\label{tab:blackjax-runtime}
\end{table*}

Moving beyond CPU execution, the GPU implementation exhibits effectively constant run-time as model dimensionality increases, $\mathcal{O}(1)$, indicating near-perfect parallelisation of the likelihood evaluation with increasing parameters. Although this behaviour will ultimately be limited by available device memory, these results demonstrate that even for models with region counts well beyond those required for realistic analyses, the memory capacity of a modern accelerator such as the NVIDIA A100 is sufficient.

For scaling with data volume, the effect of JIT execution is not shown explicitly at the data scales investigated (up to 2000 spectra, representative of a typical six-month REACH telescope observing window) as the computational overhead of non-compiled CPU execution renders such benchmarking infeasible. It is important to note, however, that this represents how joint, time-resolved fits at these scales was entirely impractical within the traditional CPU-based pipeline and therefore the current framework is essential for processing the full volume of observational data. We therefore focus on the comparative scaling behaviour of the JIT-compiled CPU and GPU implementations using a constant 10-region model, as shown in Figure~\ref{fig:likelihood_timing_data}. Once again we show that the sequential processing nature of CPUs leads to increased runtime with data volume in comparison to $\mathcal{O}(1)$ scaling on a GPU. 

One noticeable trend in the GPU runtime is a discontinuity at 108 time samples, resulting in a $\approx 20$ \%  reduction in runtime. We attribute this to the hardware specifications of the NVIDIA A100 (80GB), which features 108 Streaming Multiprocessors (SMs). Therefore as the operations achieve sufficient occupancy to allow for  Tensor Core acceleration, a transition in the underlying execution strategy occurs, moving the workload from standard Single Instruction, Multiple Thread (SIMT) execution on the FP64 CUDA cores (9.7 TFLOPS) to the more efficient Single Instruction, Multiple Data (SIMD) style processing of the FP64 Tensor Cores (19.5 TFLOPS) \citep{a100_2020}.

\subsubsection{Nested Sampling Acceleration}

Finally, we benchmark the total execution time of the \texttt{BlackJAX} algorithm used in this work against a traditional CPU-based nested sampler, \texttt{PolyChord} \citep{polychord}. This comparison evaluates end-to-end pipeline performance, encompassing both the sampler's algorithmic efficiency and the accelerated likelihood evaluations. The algorithmic configurations adopted in this work to ensure highly accurate evidence convergence (Section~\ref{sec:hyperparam}) were infeasible to benchmark directly on CPU
hardware. Even the simplest case (10 regions, integrated analysis) required 6.38 hours on a 40-core Intel Ice Lake CPU node (10.64 core-days), compared to 110.3 seconds on a single A100 GPU, corresponding to an approximate financial saving of $83\times$. To enable fair like-for-like benchmarking, we therefore performed the runs in Table~\ref{tab:blackjax-runtime} with consistent configurations matching \texttt{PolyChord} defaults: \texttt{n\_live} = $25 \times N_{\mathrm{Dim}}$ and \texttt{num\_inner\_steps} = $5 \times N_{\mathrm{Dim}}$.

While we have demonstrated that likelihood evaluations can be parallelised across data volume and model complexity to achieve near-constant run-time, Table~\ref{tab:blackjax-runtime} shows that the total GPU Nested Sampling run-time is not perfectly constant. With increasing model dimensionality, this mainly reflects the growth in the number of likelihood calls required to evolve the contraction of the prior mass; for slice-sampling-based NS, this cost can scale approximately as $\mathcal{O}(N_{\mathrm{Dim}}^3)$ \citep{polychord}. Similarly, moving from integrated to time-resolved analyses increases likelihood-surface complexity, requiring more likelihood calls to satisfy the evidence-convergence criterion.

Furthermore, although the hardware used did not reach its memory limit during the benchmarking of individual likelihood calls, the concurrent launching of parallel slice sampling runs, the number of which scales with the number of parameters and the degree of parallelisation (\texttt{num\_delete} $\times$ \texttt{n\_dim}), will eventually saturate the device memory. Despite these effects, the scaling remains significantly more efficient than CPU-based methods.  As shown in \autoref{tab:blackjax-runtime}, the speed-up factor of the GPU pipeline increases rapidly with model complexity, reaching a value of 52.71 for the 30-region resolved case. This improvement effectively makes high-dimensional, time-resolved fits feasible for physically motivated 21-cm analyses such as the REACH pipeline.

Overall, under the algorithmic configurations required for this work, the GPU-accelerated pipeline completes the 1052 runs summarised in Figure~\ref{fig:mastergrid} in under two days. In contrast, extrapolating the performance of the traditional CPU-based pipeline suggests that the same suite of fits would require approximately 100 CPU core-years, corresponding to an estimated financial saving by a factor of $\sim 180$ (see the caption of Table~\ref{tab:blackjax-runtime} for benchmark hardware configurations and pricing assumptions).

\section{Region Construction}
\label{sec:regionsplittng}

The accuracy with which the diffuse Galactic and extragalactic foregrounds can be approximated using the framework introduced in Section~\ref{sec:ForwardModel} is intrinsically linked to both the number of regions adopted and the spatial logic used to define the corresponding masks.

If the partitioning is overly coarse, the assumption that extended regions of the sky can be modelled with a single spectral index or amplitude scale factor fails. Such under-parameterisation leaves systematic residuals in the modelled antenna temperature that may bias, or potentially mimic, the global 21-cm signal. Conversely, over-parameterisation through excessive subdivision incurs a prohibitive computational cost. Specifically, the number of required likelihood evaluations within slice-sampling-based NS algorithms (Section~\ref{sec:gpunestedsampling}) are highly sensitive to the dimensionality of the parameter space, in the worst case scaling as $\mathcal{O}(N_{\mathrm{Dim}}^3)$ \citep{polychord}. Model compactness is therefore a critical consideration for practical implementation, especially given that robust analyses typically require large ensembles of inference runs for evidence-driven optimisation and validation.

This section first reviews the previously adopted sky-partitioning scheme and discusses its limitations (Section~\ref{sec:oldsplitting}), before introducing the methodology developed in this work to address these shortcomings (Section~\ref{sec:regionalcdf} and Section~\ref{sec:splittingschemes}). A quantitative comparison between the two approaches is presented in Section~\ref{sec:results}.

\subsection{Traditional Partitioning and the Occam Penalty}
\label{sec:oldsplitting}

Typically, regional partitioning has relied on static, observation-independent masks defined by subdividing the spectral index map, $\beta(\theta,\phi,t)$ (Equation~\ref{eqn:spectralindexmap}), into $N_{\beta}$ regions using either (i) `linear splitting', in which $\beta$ is divided into $N_{\beta}$ uniform intervals of equal width, or (ii) `percentile splitting', in which the bin edges are set by the percentiles of the global $\beta$ distribution such that each region contains approximately equal numbers of pixels.

One of the primary advantages of the Bayesian evidence $\mathcal{Z}$ is its intrinsic penalisation of additional model complexity that does not substantially improve the fit, a manifestation of Occam's razor \citep{MacKay_1992}. However, in both the linear and percentile splitting schemes, increasing $N_{\beta}$ to $N_{\beta}+1$ does not simply add a new dimension to the existing posterior, it redefines all region boundaries across the sky and hence the entire parameter space. 

Because the previous model configuration is not preserved as a nested subset of the new one, there is no guarantee that the $N_{\beta}+1$ case will recover or exceed the maximum likelihood, $\mathcal{L}_{\text{max}}$, of its predecessor. This inconsistency obscures the Occam’s penalty, as fluctuations in evidence are driven simultaneously by changing spatial definitions and increasing model flexibility. Furthermore, the loss of a monotonically increasing $\mathcal{L}_{\text{max}}$, a robust indicator of algorithmic convergence as model complexity grows, removes a critical diagnostic for ensuring sufficient exploration of the increasingly high-dimensional likelihood surface.

Furthermore, because the sky brightness is spatially non-uniform and modulated by the antenna beam, regions defined under these methods contribute unevenly to the observed antenna temperature $T_{\mathrm{data}}(\nu)$. This lack of observational awareness leads to an inefficient allocation of degrees of freedom: parameters associated with regions of low beam-convolved sky brightness remain prior-dominated, inflating the dimensionality of the model space without significant improvement to  the suppression of foreground systematics below that required for accurate signal recovery.

Tackling these limitations therefore requires addressing two primary challenges. First, considering the regional sky contributions to the observed antenna temperature, and second, ensuring that successive refinements maintain a nested partitioning structure. We propose a two-stage methodology to achieve this: the definition of an importance-weighted representation of the spectral index distribution (Section~\ref{sec:regionalcdf}), followed by the application of recursive algorithms to subdivide that distribution (Section~\ref{sec:splittingschemes}).

\begin{figure}
    \centering
    \includegraphics[width=\columnwidth]{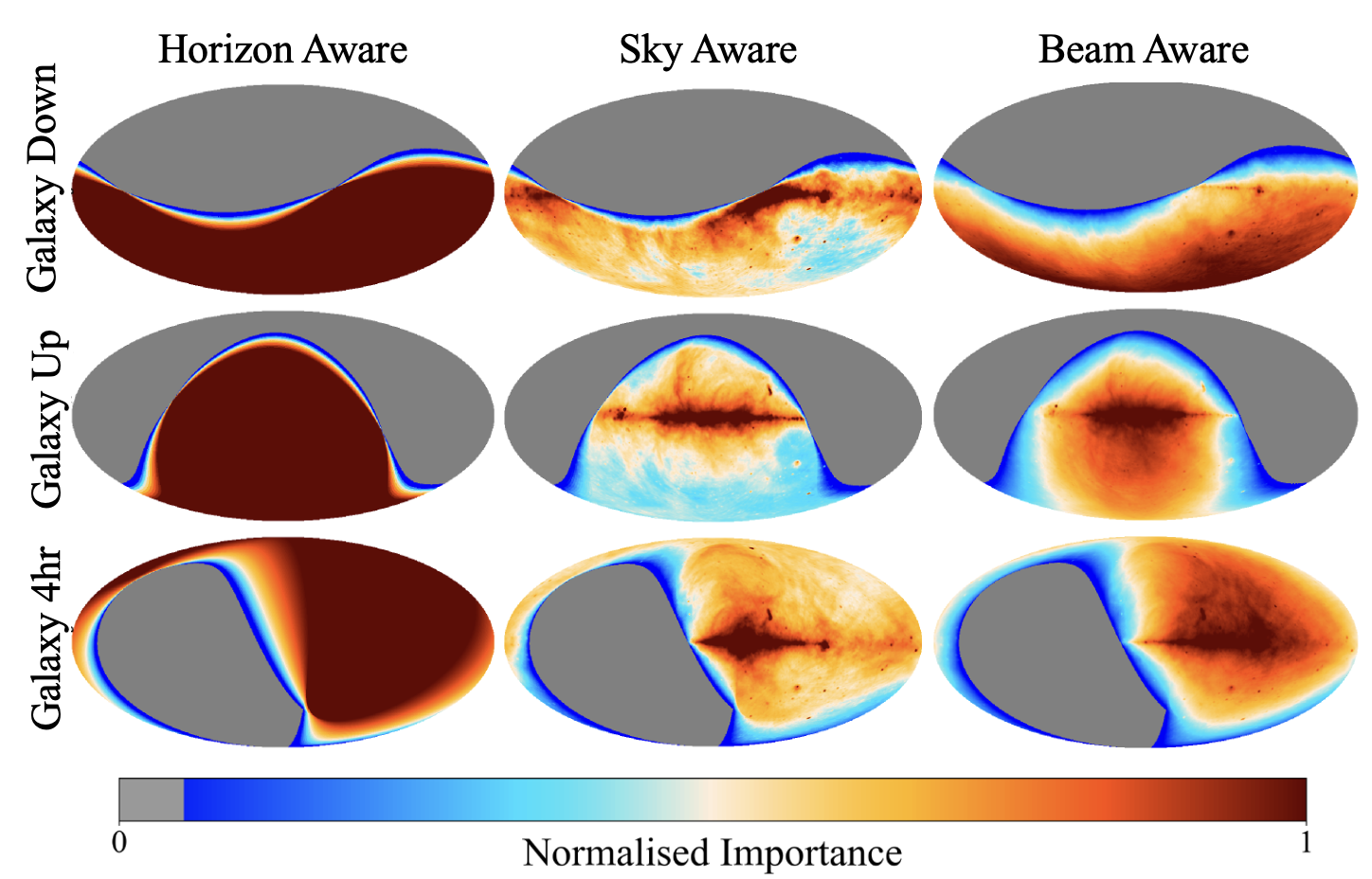}
    \caption{Time-integrated importance maps in Galactic coordinates used to construct sky regions. Each row corresponds to an observing window: Galaxy Down (1hr), Galaxy Up (1hr) and a 4hr Galaxy integration, and each column shows a different weighting: "horizon aware” (visibility-only), "sky aware” (sky-brightness-weighted), and “beam aware” (beam-convolved sky-brightness-weighted).  Colours indicate relative importance, with higher-importance regions shown in red, lower-importance regions in blue, and grey indicating zero weight. The beam-aware maps concentrate weight where the instrument is both sensitive and the sky is bright, producing observation-dependent importance structures that motivate adaptive region definitions for foreground parametrisation.}
    \label{fig:importance_grids}
\end{figure}

\subsection{Spectral Index Importance Weighting}
\label{sec:regionalcdf}

\begin{figure*}
    \centering
    \includegraphics[width=\textwidth]{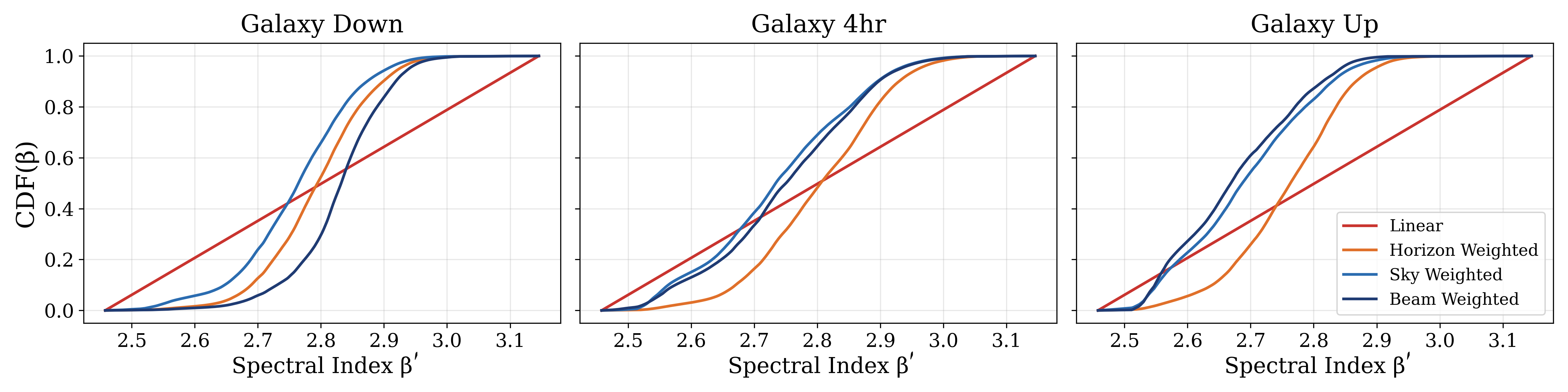}
    \caption{Spectral-index cumulative distribution functions (CDFs) constructed from different sky-weighting schemes. Each panel shows the empirical CDF, $C(\beta')$, of the diffuse-foreground spectral index threshold $\beta'$ for a given observing window: Galaxy Down (1hr), Galaxy Up (1hr) and a 4hr Galaxy integration. The unweighted linear mapping is uniform in $\beta'$, while the horizon-, sky-, and beam-weighted CDFs are constructed from the base spectral index map weighted by the corresponding time-integrated pixel importances shown in Figure~\ref{fig:importance_grids}.}
    \label{fig:cdf}
\end{figure*}

To better inform the construction of the foreground model for a given observation, we introduce an importance weighting map, $W(\theta, \phi, \nu, t)$. This quantifies the fractional relevance of each spatial coordinate to the total measured data, ensuring that the model's degrees of freedom are allocated where they have the most weight.

In the following, we first outline a series of increasingly sophisticated weighting schemes, progressing from simple observability constraints to complex instrument-aware sensitivity. We then describe how these high-dimensional maps are compressed into a one-dimensional cumulative distribution, $C(\beta')$, which serves as the foundation for defining region masks downstream. 

\subsubsection{Horizon-Aware Weighting}
The most fundamental weighting scheme considers the observability of the sky overhead given the constraints introduced by the local environment. By incorporating the static binary horizon mask $H(\theta, \phi)$ \citep[for full details see][]{Pattison_2023}, this scheme identifies the subset of the spectral index distribution that is physically observable to the antenna at any given time. It weights the importance of each coordinate accordingly, ensuring the model is grounded by the observation's field of view across its Local Sidereal Time (LST) range:
\begin{equation}
    W_{\mathrm{horizon}}(\theta, \phi) = \frac{H(\theta, \phi)}{\int_{4\pi} H(\theta', \phi') \, \mathrm{d}\Omega'} \, .
\end{equation}

\noindent This effect is illustrated for all three observation windows in the left column of Figure~\ref{fig:importance_grids}, revealing the observation-dependent nature of the importance assigned to given segments of the galaxy. Specifically, regions near the edges of the observable track exhibit a tapering of importance, reflecting the shorter duration for which they remain within the field of view as the sky rotates across the observing period.

\subsubsection{Sky-Brightness-Aware Weighting}

The horizon-aware weighting can be further refined by incorporating the distribution of celestial power, $T_{\mathrm{sky}}(\theta, \phi, \nu, t)$ (as defined in Equation~\ref{eqn:skyapproximation}). This scheme recognises that high-intensity regions, such as the Galactic centre, exert a disproportionate influence on the total antenna temperature. By scaling the significance of each coordinate relative to its brightness, the resulting distribution ensures that the forward model's spatial resolution is concentrated on the regions that dominate the incident sky brightness temperature:

\begin{equation}
    W_{\mathrm{sky}}(\theta, \phi, \nu, t) = \frac{T_{\mathrm{sky}}(\theta, \phi, \nu, t) H(\theta, \phi)}{\int_{4\pi} T_{\mathrm{sky}}(\theta', \phi', \nu, t) H(\theta', \phi') \, \mathrm{d}\Omega'} \,.
\end{equation}

\noindent The increased importance assigned to the Galactic plane is readily apparent in the comparison between the left and middle columns of Figure~\ref{fig:importance_grids}, where the brightness-aware weighting refocuses the model's priority.

\subsubsection{Beam-Aware Weighting}
While previous schemes assume uniform sensitivity across the visible sky, the final refinement incorporates the antenna beam, $D(\theta, \phi, \nu)$. Through calculating the convolution of the beam pattern with sky brightness across all LSTs, this weighting accounts for the spatial and frequency-dependent sensitivity of the instrument. Consequently, this provides the most accurate measure of the spatial foreground contribution to the observational data:
\begin{equation}
\label{eqn:w_beam_dynamic}
\begin{split}
    W_{\mathrm{beam}}(\theta, \phi, & \nu, t)  = \\ 
    & \frac{T_{\mathrm{sky}}(\theta, \phi, \nu, t) D(\theta, \phi, \nu) H(\theta, \phi)}{\int_{4\pi} T_{\mathrm{sky}}(\theta', \phi', \nu, t) D(\theta', \phi', \nu) H(\theta', \phi') \, \mathrm{d}\Omega'}  \, .
\end{split}
\end{equation}

\noindent The refinement introduced by the beam-weighted scheme is illustrated by comparing the middle and right-hand columns of Figure~\ref{fig:importance_grids}, revealing a nuanced interplay between celestial power and instrumental sensitivity. For the `Galaxy Up' case, importance is further concentrated at the zenith, where the Galactic plane aligns with the primary beam. In contrast, for `Galaxy Down', while the high-zenith Galactic regions remain a dominant power source, their influence is mitigated by the reduced beam sensitivity at those angles, as characterized by the chromatic beam pattern in Figure~\ref{fig:chromatic_beam}.

It should be noted that both the derived beam- and sky-weighting are intrinsically tied to the accuracy of the sky model used to derive $T_{\text{sky}}$ as defined in Equation~\ref{eqn:simantenna} or similiarly the accuracy of the adopted beam model, $D(\theta,\phi,\nu)$ under the precision tolerances of modern electromagnetic simulations. However, as this weighting is ultimately summarised by a one-dimensional distribution (Equation~\ref{eqn:weighted_cdf}), the resulting region partitions are relatively robust to minor local inaccuracies. This is evidenced by the relative behaviour of the weighting schemes: while beam- and sky-aware refinements provide measurable benefits, the dominant step-change is the transition from the observation-independent linear split to any sky-aware nested formulation. Differences among the advanced weighting variants are therefore second-order, indicating that moderate prior-map or beam-model inaccuracies are unlikely to alter the principal partitioning gains (see Section~\ref{sec:results}).

\subsubsection{The Importance-Weighted Distribution} 

To transform the spatial importance maps into a form suitable for partitioning, we project them onto the spectral-index coordinate. The resulting Cumulative Distribution Function (CDF) provides a principled basis for region definition, with boundaries informed by the relative observational contribution of different spectral index ranges. We define $C(\beta')$ as:

\begin{equation} 
\label{eqn:weighted_cdf} 
C(\beta') = \int_{T} \int_{B} \int_{4\pi} W(\theta, \phi,\nu, t) \, \mathbbm{1}_{\beta'}\!\big(\beta(\theta,\phi,t)\big) \, \mathrm{d}\Omega \, \mathrm{d}\nu \,  \mathrm{d}t \, , \end{equation}

\noindent where $T$ and $B$ denote the LST range and frequency band of the observation, respectively, and $\mathbbm{1}_{\beta'}(y)=1$ if $y\le \beta'$ and $0$ otherwise. The integrations over sky position, frequency and time therefore accumulate the importance assigned to all observed elements whose spectral index lies below the threshold $\beta'$. The impact of incorporating increasing levels of physical detail is summarised by the variations in the resulting CDFs, detailed in Figure~\ref{fig:cdf}. Specifically, a shift in the distribution toward lower spectral indices signifies an increased importance assigned to the Galactic centre, whereas a shift toward higher indices represents a greater observational contribution from the Galactic poles. We note that, while the CDFs constructed in this work are physically motivated, the region-construction framework is not restricted to this choice and may be generalised to alternative parametric forms, enabling more flexible optimisation of the region definitions.

\begin{figure}
    \centering
    \includegraphics[width=\columnwidth, trim=0 0 0 0, clip]{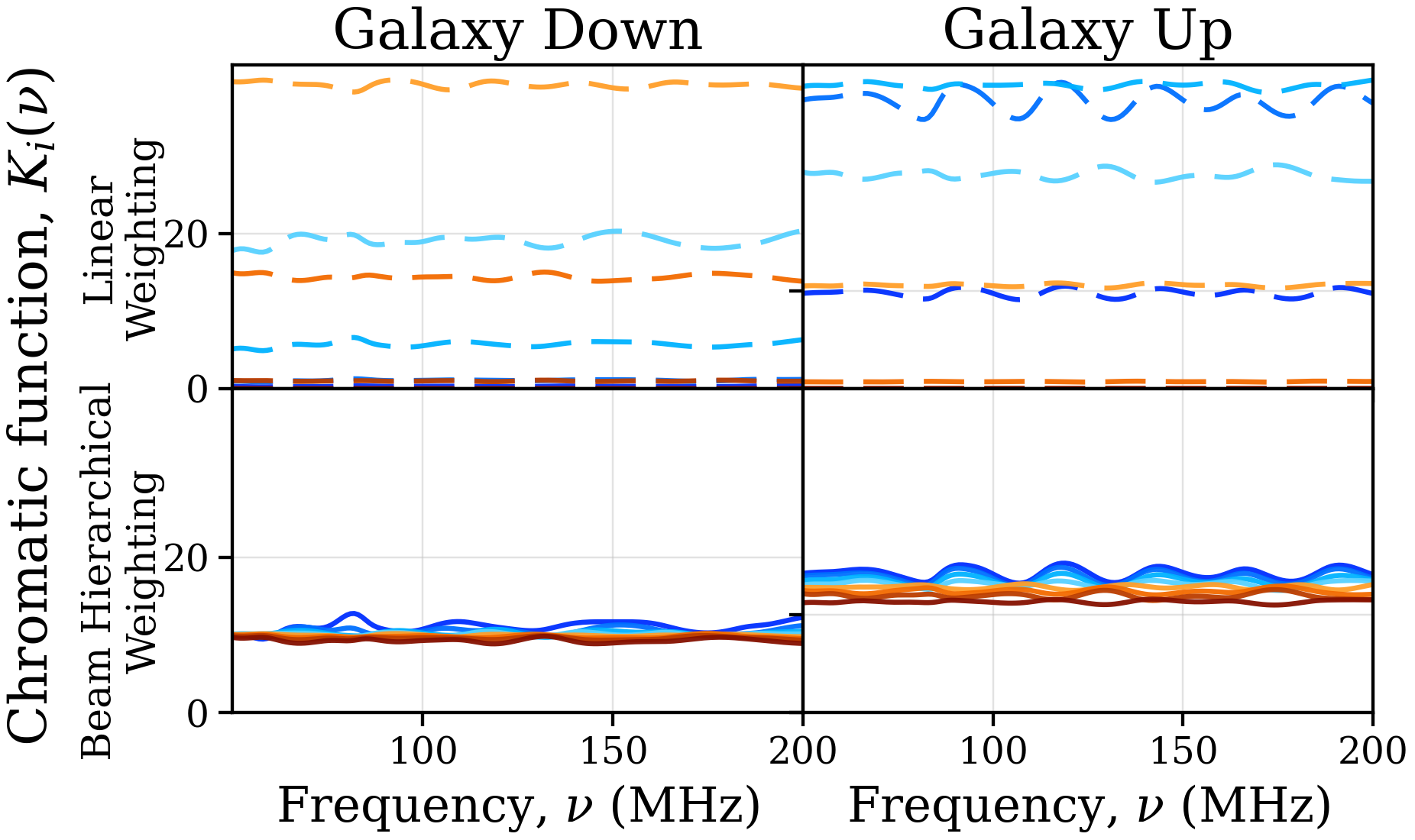}
    \caption{A comparison of chromatic functions, $K_i(\nu)$, for $N_{\rm reg}=8$ regions defined using observation-independent linear weighting and observation-dependent beam-hierarchical weighting schemes. The columns show the Galaxy Down and Galaxy Up observing windows. The top row shows the functions for the linear split, while the bottom row shows the corresponding beam-hierarchical construction. Each coloured curve represents an individual sky region.}
    \label{fig:chromatic_range}
\end{figure}

\begin{figure*}
    \includegraphics[width=\textwidth]{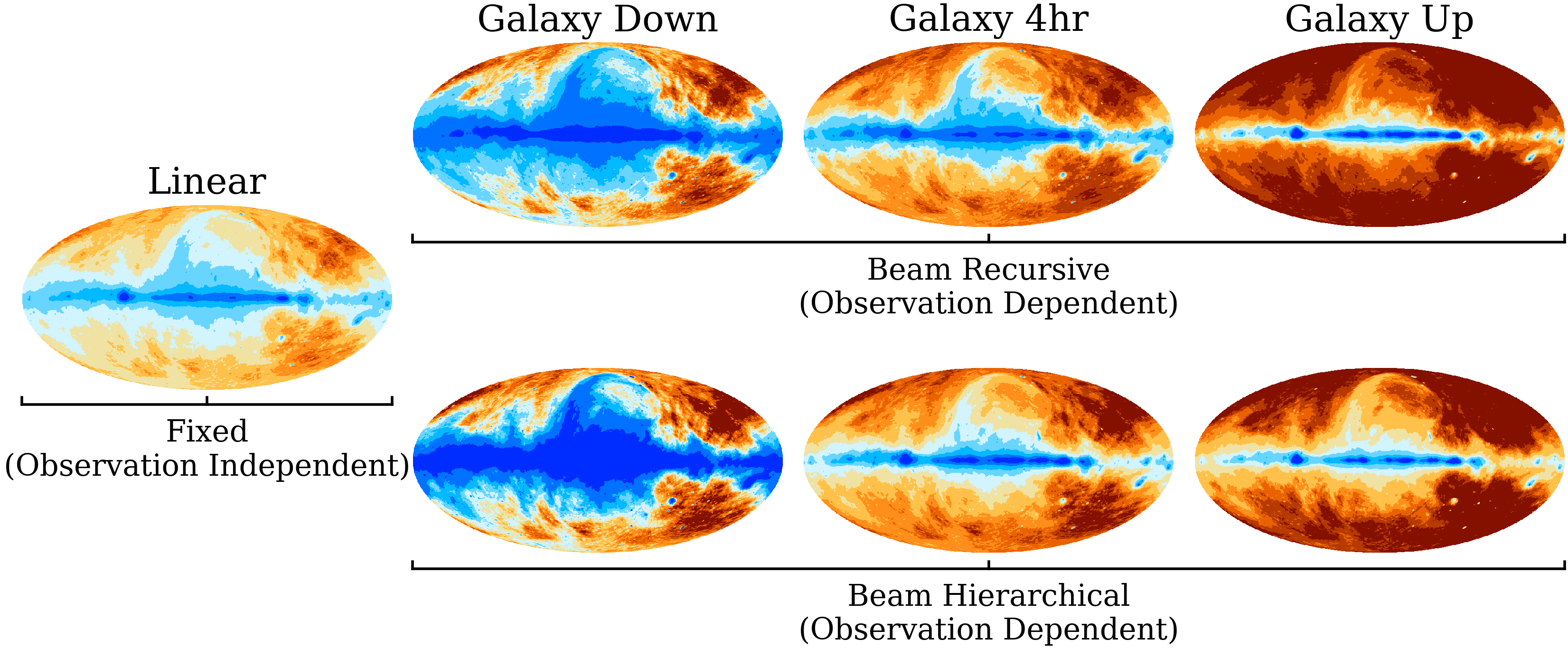}
    \caption{Comparison of discrete sky splitting masks $M_{\beta,j}$ for $N_{\rm reg} = 11$ regions in Galactic coordinates. The columns compare an observation-independent linear splitting baseline with observation-dependent beam-weighted schemes for three observing windows: Galaxy Down (1 hr), a 4-hr Galaxy integration, and Galaxy Up (1 hr). The top row shows the beam-recursive partitioning scheme, while the bottom row shows the beam-hierarchical scheme constructed from the same importance-weighted CDFs. Each colour represents a unique binary mask $M_{\beta,j}$, using a discrete version of the Planck-like colour scale.}
    \label{fig:Region_definition}
\end{figure*}

\subsection{Algorithmic Partitioning Schemes}
\label{sec:splittingschemes}

Given the importance-weighted CDF established in Section~\ref{sec:regionalcdf}, discrete sky regions are constructed by partitioning the resulting one-dimensional distribution optimally. We propose two distinct algorithmic frameworks, \textit{Hierarchical Partitioning} (Section~\ref{sec:hierpart}) and \textit{Recursive Partitioning} (Section~\ref{sec:rescpart}), differing in their strategy of allocating parameters across the spectral index domain. Crucially, both methodologies enforce a strictly nested model hierarchy enabling better informed model selection.

\begin{algorithm}
\caption{Hierarchical Partitioning}\label{alg:method1}
\begin{algorithmic}[1]
\State \textbf{Input:} Total regions $N_{\mathrm{total}}$, weighted CDF $C(\beta')$
\State \textbf{Step 1: Base Partition Construction}
\State $N_{\mathrm{base}} \gets 2^{\lfloor \log_2(N_{\mathrm{total}}) \rfloor}$ \Comment{Largest power of 2 $\leq N_{\mathrm{total}}$}
\State $\mathcal{R} \gets$ Partition $C(\beta)$ into $N_{\mathrm{base}}$ equal-mass regions
\State \textbf{Step 2: Residual Resolution Enhancement}
\State $N_{\mathrm{rem}} \gets N_{\mathrm{total}} - N_{\mathrm{base}}$
\While{$N_{\mathrm{rem}} > 0$}
    \State \textbf{Target:} Identify $r^* = [\beta_a, \beta_b] \in \mathcal{R}$ with max width:
    \Statex \hspace{\algorithmicindent} $\Delta \beta_{r^*} = \beta_b - \beta_a$
    \State \textbf{Split:} Bisect $r^*$ at its importance midpoint:
    \Statex \hspace{\algorithmicindent} $C_m = \frac{1}{2}\big(C(\beta_a) + C(\beta_b)\big)$
    \Statex \hspace{\algorithmicindent} $\beta_m = C^{-1}(C_m)$
    \State \textbf{Update:} $\mathcal{R} \gets (\mathcal{R} \setminus \{r^*\}) \cup \{[\beta_a, \beta_m], [\beta_m, \beta_b]\}$
    \State $N_{\mathrm{rem}} \gets N_{\mathrm{rem}} - 1$
\EndWhile
\State \textbf{Return:} $\mathcal{R}$
\end{algorithmic}
\end{algorithm}

\begin{algorithm}
\caption{Recursive Partitioning}\label{alg:method2}
\begin{algorithmic}[1]
\State \textbf{Input:} Total regions $N_{\mathrm{total}}$, weighted CDF $C(\beta')$
\State \textbf{Step 1: Initialisation}
\State $\mathcal{R} \gets \{[\beta_{\min}, \beta_{\max}]\}$ \Comment{Start with a single region}
\State \textbf{Step 2: Iterative Mass-Targeted Splitting}
\While{$|\mathcal{R}| < N_{\mathrm{total}}$}
    \State \textbf{Measure:} For each $r \in \mathcal{R}$, calculate $m_r = \int_{r} \mathrm{d}C(\beta)$
    \State \textbf{Target:} Identify $r^* = [\beta_a, \beta_b] \in \mathcal{R}$ with max mass:
    \Statex \hspace{\algorithmicindent} $m_{r^*} = \max(\{m_r \mid r \in \mathcal{R}\})$
    \State \textbf{Split:} Bisect $r^*$ at its physical midpoint:
    \Statex \hspace{\algorithmicindent} $\beta_m = \frac{1}{2}(\beta_a + \beta_b)$
    \State \textbf{Update:} $\mathcal{R} \gets (\mathcal{R} \setminus \{r^*\}) \cup \{[\beta_a, \beta_m], [\beta_m, \beta_b]\}$
\EndWhile
\State \textbf{Return:} $\mathcal{R}$
\end{algorithmic}
\end{algorithm}

\subsubsection{Hierarchical Partitioning} 
\label{sec:hierpart}

The Hierarchical scheme (Algorithm ~\ref{alg:method1}) first establishes a base partition of $N_{\mathrm{base}} = 2^{\lfloor \log_2 N_{\mathrm{total}} \rfloor}$ regions, the largest power of two less than or equal to the desired total resolution. These initial regions are defined by equal-mass intervals in the cumulative distribution, such that each region represents an equivalent fraction of the total importance, $\Delta C = 1/N_{\mathrm{base}}$.

When the desired resolution $N_{\mathrm{total}}$ is not a power of two, the remaining $N_{\mathrm{rem}}$ degrees of freedom are allocated through a refinement step. The algorithm identifies existing regions with the largest width in spectral index space, $\Delta\beta = \beta_b - \beta_a$, and bisects them. Crucially, this bisection occurs at the `importance midpoint', $C_m = \frac{1}{2}\big(C(\beta_a) + C(\beta_b)\big)$, where $\beta_m = C^{-1}(C_m)$ is the spectral index value corresponding to that cumulative mass.

\subsubsection{Recursive Partitioning}
\label{sec:rescpart}

In contrast, the Recursive scheme (Algorithm~\ref{alg:method2}) prioritises regions of maximum importance density through iterative refinement. Initialised with a single global region covering $[\beta_{\min}, \beta_{\max}]$, the algorithm recursively identifies the region $r^*$ that encapsulates the highest total importance mass, $m_{r^*}$, defined as the integral of the weighted distribution across that interval: $m_r = \int_{r} \mathrm{d}C(\beta)$. Once the highest-mass region is identified, it is bisected at its spectral index midpoint, $\beta_m = \frac{1}{2}(\beta_a + \beta_b)$. This subdivision is repeated until the total number of regions reaches $N_{\mathrm{total}}$.

\subsubsection{Comparative Performance}

The key gain from the observation-dependent schemes is common to both the Hierarchical and Recursive algorithms. In each case, the importance-weighted CDF allocates foreground parameters according to a region's contribution to the measured antenna temperature, rather than according to an observation-independent spectral-index range. This shared effect is illustrated in Figure~\ref{fig:chromatic_range} for the beam-hierarchical case: relative to the linear split, the beam-aware weighting compresses the dynamic range of the regional chromatic functions. This produces regions that are more similarly constrained by the data, directly addressing the inefficiencies of observation-independent linear splitting discussed in Section~\ref{sec:oldsplitting}. Within this common framework, the algorithms differ in their order of prioritisation. Hierarchical Partitioning first ensures that each region has uniform importance mass, while Recursive Partitioning prioritises addressing the assumption that a given region can be defined by a single spectral index while balancing the requirement for higher parameter density where the foregrounds are most dominant.

In practice, while both algorithms showed advantages over the linear splitting method, the Recursive Partitioning framework offered a more efficient optimisation of the parameter space compared to the hierarchical alternative thus providing a superior ability to suppress systematic residuals while maintaining model compactness. Given this balance of accuracy and computational speed, it is the focus of the results presented hereafter. The impact of these observationally-dependent schemes on the resultant foreground masks, $M_{\beta,j}$, is illustrated in Figure~\ref{fig:Region_definition}. The dominant visual change is the redistribution of regions relative to the observation-independent linear baseline, driven by the Galactic orientation and the beam weighting. The differences between the beam-recursive and beam-hierarchical masks are more subtle, reflecting the fact that both are derived from the same importance-weighted CDF but apply different rules for assigning its mass to discrete sky regions.

\section{Statistical Validation}
\label{sec:statisticalvalidation}
\begin{figure*}
    \centering
    \includegraphics[width=\textwidth, trim={0 0 0 0}, clip]{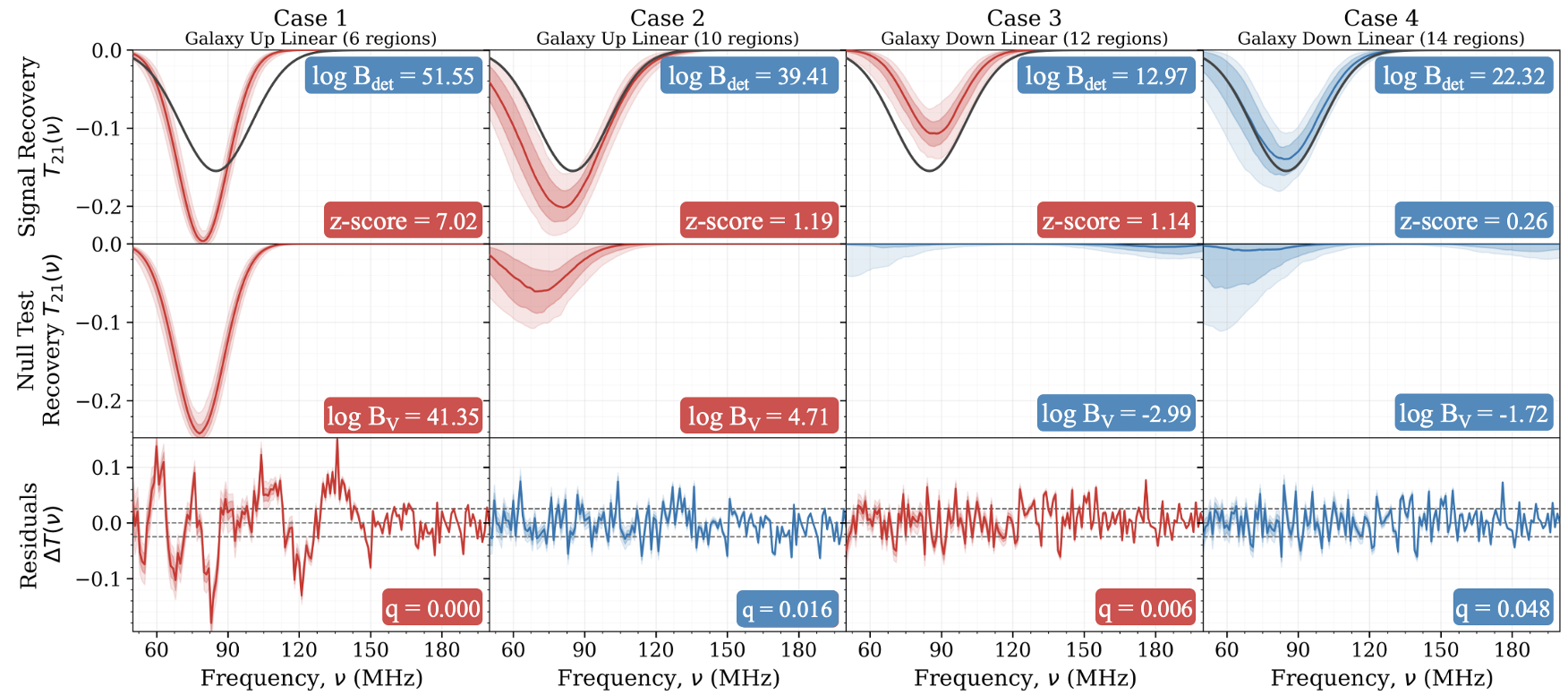}
	    \caption{Illustration of the validation framework applied to representative signal-recovery scenarios. Each column corresponds to a distinct inference outcome, all of which exhibit a decisive statistical preference for the inclusion of a 21-cm signal model. Rows show (top to bottom) the recovered signal posterior for signal-injected data, the corresponding posterior obtained from the null test (no signal injected), and the residuals of the overall best-fitting model. Shaded bands indicate the posterior mean and $1\sigma$ and $2\sigma$ credible intervals. Blue and red curves denote configurations that pass or fail the respective validation criteria. Boxed annotations report the detection Bayes factor $\log B_{\mathrm{det}}$, the signal-recovery $Z$-score, and the null-test evidence ratio $\ln B_{\mathrm{val}}$, highlighting cases in which statistically favoured detections nonetheless correspond to biased or unreliable signal recovery.
	    }
    \label{fig:validationex}
\end{figure*}

While evaluating the significance of a potential detection via the Bayes factor is an essential step toward statistical rigor, the extreme sensitivity of signal recovery to foreground mismodelling means that, in isolation, it is insufficient to guarantee a physical detection. Given the aim of this work is to optimise the parametrisation of foreground models, it is essential to benchmark these refinements against a robust Bayesian validation framework. To address this, we adopt the methodology presented in Pattison et al. (2026, in prep), which describes the integration of the BaNTER validation framework \citep{Sims_2025} into physically motivated Global-21cm analysis pipelines. In Section~\ref{sec:validtionsfailure}, we discuss the specific failure modes of pure evidence-based comparisons, before defining the two key validation metrics used to identify and flag incorrect recoveries in Section~\ref{sec:validationsolutions}. 

\subsection{Failure Modes and Model Degeneracy} 
\label{sec:validtionsfailure}

Although the detection Bayes factor, $B_{\mathrm{det}}$, assesses whether the increased flexibility of a joint foreground-plus-signal model, $\mathbf{M}_{\mathrm{FG+21}}$, yields a better statistical fit than a foreground-only model, $\mathbf{M}_{\mathrm{FG}}$, it represents a blind comparison and is agnostic to the physical origin of that improvement. Due to inherent model-level degeneracies \citep{Sims_2025b, liu_2025} between the chromatic structure introduced by diffuse foregrounds and the global 21-cm signal, this metric cannot distinguish between cases where the signal recovery is reliable, or whether the 21-cm signal model is compensating for inaccuracies within the foreground model and hence the recovery is biased.

As an illustration of the range of possible inference outcomes, Figure~\ref{fig:validationex} presents four representative cases of signal recovery in which the inclusion of a 21-cm signal model is decisively favoured according to the criteria in \autoref{tab:jeffreys}. Despite this strong statistical preference, only one case yields an accurate recovery of the injected signal parameters. Throughout this work, we quantify the accuracy of signal recovery using an uncertainty-aware metric defined in Equation~\ref{eqn:zscore}, hereafter referred to as the Z-score, and classify recoveries with $Z < 1$ as accurate,

\begin{equation}
\label{eqn:zscore}
Z
=
\frac{1}{N_{\mathrm{sig}}}
\sqrt{
\sum_{i=1}^{N_{\mathrm{sig}}}
\frac{(\mu_i - \theta_{i,\text{true}})^2}{\hat{\sigma}_i^2}
},
\end{equation}

\noindent where $\mu_i$ is the posterior mean, $\theta_{i,\mathrm{true}}$ is the true injected value, and $\hat{\sigma}_i$ is the posterior standard deviation for each of the $N_{\mathrm{sig}}$ signal parameters. The summation is therefore restricted to the cosmological signal parameters and does not include foreground or other nuisance parameters. Effectively, this $Z$-score represents the average distance in signal-parameter space between the recovered mean and the true value, expressed as a multiple of the posterior standard deviation, providing a one-dimensional measure of how consistent the recovered 21-cm signal is with the ground truth.

We note that this mean-based summary is used only as a compact diagnostic for cross-referencing signal recovery against the validation metrics, and would be inappropriate if the signal posterior were strongly multimodal. The relevant posterior distributions were inspected for the configurations considered here, and were found to be sufficiently unimodal for this metric to provide a representative measure of recovery accuracy.

\subsection{Validation Checks}
\label{sec:validationsolutions}

To guard against the failure modes described above, we employ a two-stage validation strategy that jointly assesses signal–systematic degeneracies through a null test (Section~\ref{sec:null}) and evaluates the overall statistical consistency of a given fit through analysing the remaining residual structure (Section~\ref{sec:residual}).

\subsubsection{Null Tests}
\label{sec:null}

The null test probes the susceptibility of a foreground model to spurious signal detection. We apply this test for a given observational period and parameterised foreground model by fitting the associated joint model, $\mathbf{M}_{\mathrm{FG+21}}$, to a validation dataset $D_{\mathrm{V}}$ that is intentionally simulated without an injected 21-cm signal. The resulting fit is then compared to a foreground-only model, $\mathbf{M}_{\mathrm{FG}}$, via the null-test evidence ratio:

\begin{equation}
\ln B_{\mathrm{val}}
=
\ln \left(
\frac{\mathcal{Z}_{\mathrm{FG+21}}^{\mathrm{v}}}
{\mathcal{Z}_{\mathrm{FG}}^{\mathrm{v}}}
\right),
\end{equation}

\noindent where $\mathcal{Z}_{\mathrm{FG+21}}^{\mathrm{v}}$ and $\mathcal{Z}_{\mathrm{FG}}^{\mathrm{v}}$ denote the Bayesian evidences of the respective models. Due to the lack of signal within the validation data, any Bayes ratio favoring the composite model indicates that the signal component is compensating for residual foreground structure and therefore identifies foreground models that are insufficiently expressive and pose a high risk of biased detections when applied to observational data. Any configuration with $\ln B_{\mathrm{val}} \geq 0$ is thus flagged accordingly.

\subsubsection{Accuracy Criterion}
\label{sec:residual}

While the null test identifies problematic model configurations prior to their application to observational data, it does not assess whether an individual fit leaves residuals that are statistically consistent with instrumental noise. To do this, we compare the median a posteriori likelihood of a given posterior distribution, $\overline{\mathcal{L}}_i$, to the likelihood distribution expected if the data were described perfectly by the model up to random noise fluctuations. This reference distribution, denoted $\mathcal{L}_{\mathrm{noise}}$, is constructed by evaluating the likelihood using multiple realisations of the assumed noise model, taken in this work to be uncorrelated Gaussian noise with an amplitude of 25 mK.

The comparison is quantified by computing the fraction of the $\mathcal{L}_{\mathrm{noise}}$ distribution that yields likelihood values less than or equal to $\overline{\mathcal{L}}_i$:

\begin{equation}
q_i = \mathbb{P}\!\left( \mathcal{L}_{\mathrm{noise}} \le \overline{\mathcal{L}}_i \right).
\end{equation}

\noindent This represents the lower-tail probability of the fitted residual likelihood under the ideal noise distribution. Small values of $q_i$ indicate that the fitted residuals have an anomalously low likelihood compared with noise-only realisations, signalling coherent residual structure not captured by the model. We therefore classify a fit, and hence the corresponding model, as statistically consistent if it does not fall within the lower $q_{\mathrm{threshold}}$ tail of the ideal noise distribution, corresponding to $q_i \ge q_{\mathrm{threshold}}$. Throughout this work we adopt $q_{\mathrm{threshold}} = 0.01$. Fits failing this criterion are flagged as containing residual systematic structure, indicative of foreground mismodelling or signal–systematic degeneracy.

We note that for the specific case of independent Gaussian noise considered here, a simpler distributional diagnostic, such as a Kolmogorov--Smirnov (KS) test, could be used to flag non-Gaussian residuals \citep[e.g.][]{Hibbard_2023}. However, such tests primarily probe consistency through the marginal one-point distribution of the residuals. By formulating the validation criterion in terms of the residual likelihood, the same construction remains applicable if the likelihood is extended to include more structured, non-Gaussian radiometric noise models \citep{Scheutwinkel_2022a, Scheutwinkel_2022b}.

The requirement for both validation metrics to be applied in parallel is illustrated in Figure~\ref{fig:validationex}. While all cases shown strongly support the inclusion of a signal model (with $\ln B_{\mathrm{det}} \gg 0$),  validation identifies models that yield biased signal recovery (Cases 1--3). Specifically, Case 2 demonstrates a failure mode where the model accurately fits the data in aggregate, but only through biased component fits; the signal model suppresses foreground residuals enough to appear noise-like ($q > q_{\mathrm{threshold}}$), yet the null test successfully identifies this underlying degeneracy (see category II models in \citealt{Sims_2025}). Conversely, Case 3 illustrates a positive systematic not captured by either the foreground or signal models; while it passes the null test, the residual consistency criterion correctly flags the residuals as non-Gaussian ($q < q_{\mathrm{threshold}}$). Ultimately, passing both tests is found to be a reliable predictor of models that yield unbiased 21-cm signal inferences, as seen in Case 4.

\section{Results}
\label{sec:results}

\begin{figure*}
    \centering
    \includegraphics[width=\textwidth]{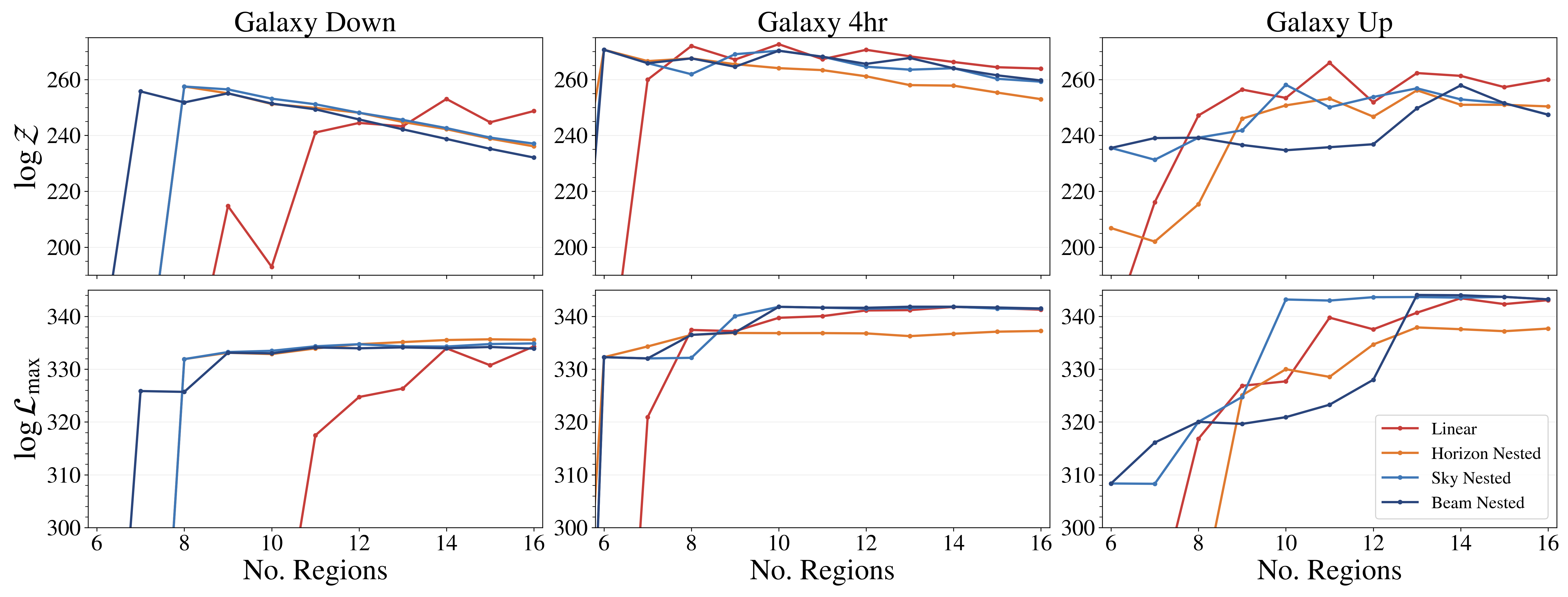}
    \caption[Bayesian evidence and maximum likelihood across sky-splitting methods]{Bayesian evidence, $\log \mathcal{Z}$ (top row) and maximum log‑likelihood, $\log \mathcal{L}_{\max}$ (bottom row) given by nested sampling versus the number of modeled foreground regions. Columns show Galaxy Down (1 hr), Galaxy Up (1 hr), and a 4‑hr Galaxy integration, while each panel compares the four splitting strategies: linear, horizon‑nested, sky‑nested, and beam‑nested.}
    \label{fig:occam}
\end{figure*}

In this section, we present the comparative results of a suite of nested sampling fits across all three observational windows, explicitly benchmarking all importance-weighted partitioning schemes against the original observation-independent linear splitting. In response to the limitations of prior region definitions discussed in Section~\ref{sec:regionsplittng}, this section is structured as follows. In Section~\ref{sec:occamresults}, we examine how nested configurations improve Bayesian model comparison. We then demonstrate the degree to which the introduction of importance-aware region definitions better constrain foreground parameters in Section~\ref{sec:foregroundresults}. Finally, in Section~\ref{sec:signalresults}, we assess whether these improvements propagate to enhanced suppression of chromatic systematics, resulting in more reliable signal recovery.

\subsection{Observing the Occam Penalty}
\label{sec:occamresults}

As discussed in Section~\ref{sec:oldsplitting}, traditional linear splitting schemes do not define a consistent model hierarchy, as the reshuffling of boundaries with increasing regions, $N_{\rm reg}$, redefines the parameter space rather than expanding it. Figure~\ref{fig:occam} illustrates the consequences of this behaviour for both the maximum log-likelihood, $\log \mathcal{L}_{\max}$, and the Bayesian evidence, $\log \mathcal{Z}$, alongside the comparative advantages of the nested schemes introduced in this work.

First, the evolution of $\log \mathcal{L}_{\max}$ under the linear splitting scheme is highly non-monotonic, demonstrating that in the absence of a nested construction, increasing model dimensionality does not guarantee an improved description of the data. In particular, successive increases in $N_{\rm reg}$ within the linear scheme can introduce region definitions that are less optimal than those of lower-dimensional models. In contrast, all nested schemes exhibit a monotonic increase in $\log \mathcal{L}_{\max}$, up to small fluctuations attributable to algorithmic convergence. This behaviour guarantees that the introduction of additional parameters can only maintain or improve the quality of the fit.

Second, when considering the global maximum likelihood attained across the full range of region counts explored ($N_{\rm reg}=6$--$16$), the physically motivated sky- and beam-weighted schemes consistently outperform the linear baseline. This demonstrates that concentrating model flexibility in observationally significant regions of the sky yields foreground models that are more representative of the underlying structure. Moreover, these improvements are achieved more efficiently, supporting the Recursive partitioning strategy introduced in Section~\ref{sec:splittingschemes}. In particular, the importance-aware schemes reach a plateau in $\log \mathcal{L}_{\max}$ at substantially lower values of $N_{\rm reg}$ than the linear scheme requires to attain a comparable quality of fit.

This effect is most pronounced in the Galaxy Down observing window, where the likelihood saturates at $N_{\rm reg}\simeq8$-$9$, while the linear scheme requires 14 or 16 regions to reach a similar level of performance. Although this trend is present across all observational windows, it is least pronounced for the Galaxy Up case, consistent with the strong Galactic emission and short integration times that necessitate higher model dimensionality for adequate resolution.

Finally, the nested constructions enable a much clearer identification of the Occam penalty through the Bayesian evidence. Because $\log \mathcal{Z}$ balances improvements in fit quality against the expansion of the prior volume, the saturation of $\log \mathcal{L}_{\max}$ in the Galaxy Down case is accompanied by a monotonic decline in $\log \mathcal{Z}$ beyond $N_{\rm reg}\simeq8$-$9$. Under the linear splitting scheme, by contrast, the evidence exhibits no coherent trend, severely limiting its utility for principled model selection. While the evidence peak is less sharply defined for the Galaxy Up window, reflecting continued competition between improving fit quality and increasing prior volume, the nested schemes nonetheless exhibit the expected behaviour once their respective likelihood maxima are reached.

It is important to note, however, that the maximisation of the Bayesian evidence alone does not guarantee sufficient foreground recovery for robust 21-cm signal inference. As discussed previously, even small residuals bias the recovered global 21-cm signal, well below the scale at which they significantly impact $\log \mathcal{Z}$, motivating the validation framework introduced in Section~\ref{sec:statisticalvalidation}.

\subsection{Foreground Recovery}
\label{sec:foregroundresults}

\begin{figure*}
    \centering
    \includegraphics[width=\textwidth, trim={80 0 0 0}, clip]{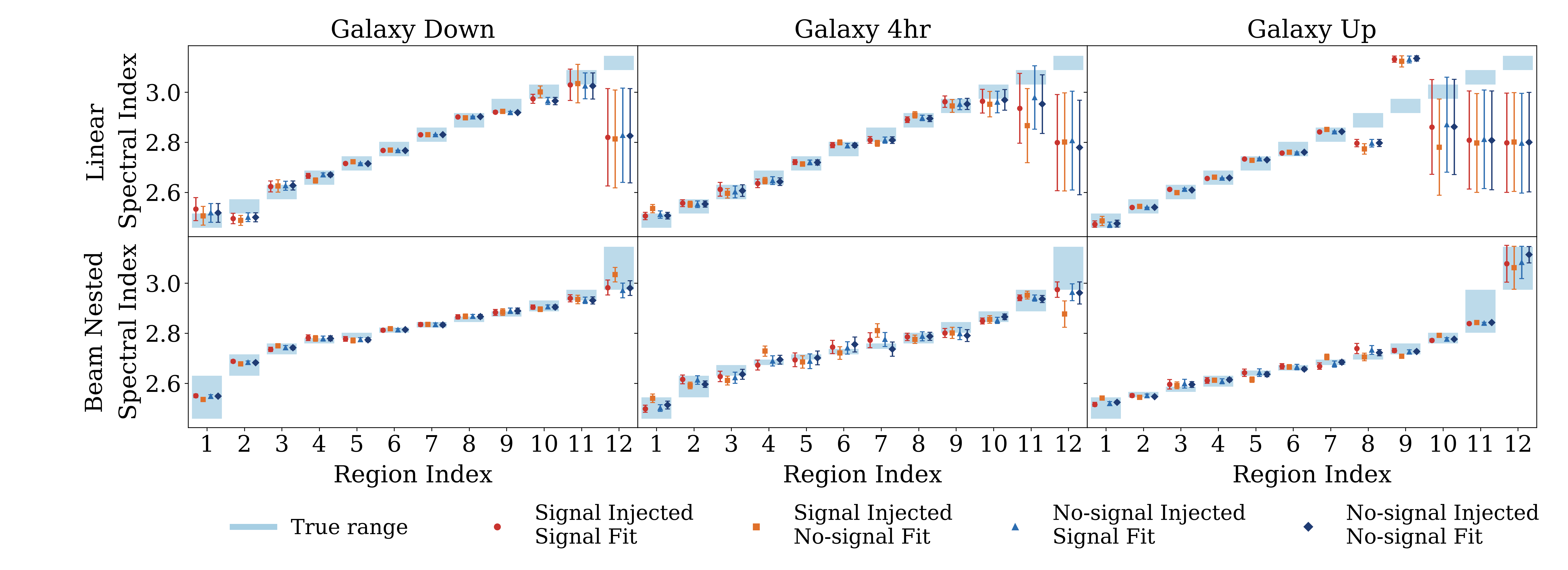}
    \caption[Foreground spectral-index recovery across sky-splitting methods]{Comparison of foreground spectral index recovery using linear vs. beam‑nested sky parameterisation schemes ($N_{\rm reg}=12$). Columns represent different observing windows (Galaxy Down, Galaxy Up, and a 4 hr Integration), while rows compare the linear splitting method (top) with the beam‑nested splitting method (bottom). In each panel, the shaded bands indicate the ground‑truth range of spectral indices ($\beta$) present within that specific region on the high‑resolution ($N_{\text{side}} = 512$) base map used for data generation. The overlaid points and $1\sigma$ error bars show recovered posterior means for the four generation/fit combinations (Signal Injected/Signal Fit, Signal Injected/No‑signal Fit, No‑signal Injected/Signal Fit, No‑signal Injected/No‑signal Fit).}
    \label{fig:specidxrecover}
\end{figure*}

Beyond the requirements of 21-cm signal extraction, the ability to update our prior state of knowledge regarding low-frequency radio maps and extrapolate them across a continuous frequency band is a valuable result in its own right. Figure~\ref{fig:specidxrecover} benchmarks the posterior spectral index recovery against the `ground truth' ranges present in the full-resolution base map used in simulations (a HEALPix map of $N_{\text{side}=512}$) for both the traditional linear splitting and the beam-nested scheme ($N_{\text{reg}=12}$).

This comparison provides a clear visualisation of the dynamic range, $\Delta \beta$, captured by each mask under varying observational conditions. In the Galaxy Up case, for instance, the adaptive masks naturally cluster around lower spectral index values, correctly prioritising the flatter-spectrum emission characteristic of the Galactic plane. Critically, we find that the recovery of the spectral index values is far more robust under the new scheme. The original linear splitting exhibits two primary shortfalls: regions associated with low beam-convolved brightness remain prior-dominated (indicated by large uncertainties and poor centring), while others are recovered with high confidence but significant bias. The Galaxy Up case being particularly inaccurate example. In contrast, the beam-nested scheme consistently centres the posterior means within the true physical range across all regions.

Furthermore, we demonstrate the robustness of this foreground recovery across various permutations of signal injection and model fitting. As expected, the recovered foreground parameters remain largely independent of the underlying 21-cm cosmology, as the signal's amplitude is orders of magnitude below the chromatic distortions. The only marginal exception is the 4-hr Galaxy integration, where the longer integration period reduces the effects of beam-sky coupling such that the presence of a 21-cm signal without modelling leads to very slight offsets in the recovered spectral indices (see Signal Injected/ No Signal Fit). While Figure~\ref{fig:specidxrecover} focuses on the 12-region case, these performance gains are consistent across all investigated values of $N_{\text{reg}}$.

\begin{figure*}
    \centering
    \includegraphics[width=0.92\textwidth, trim={20 70 100 0}, clip]{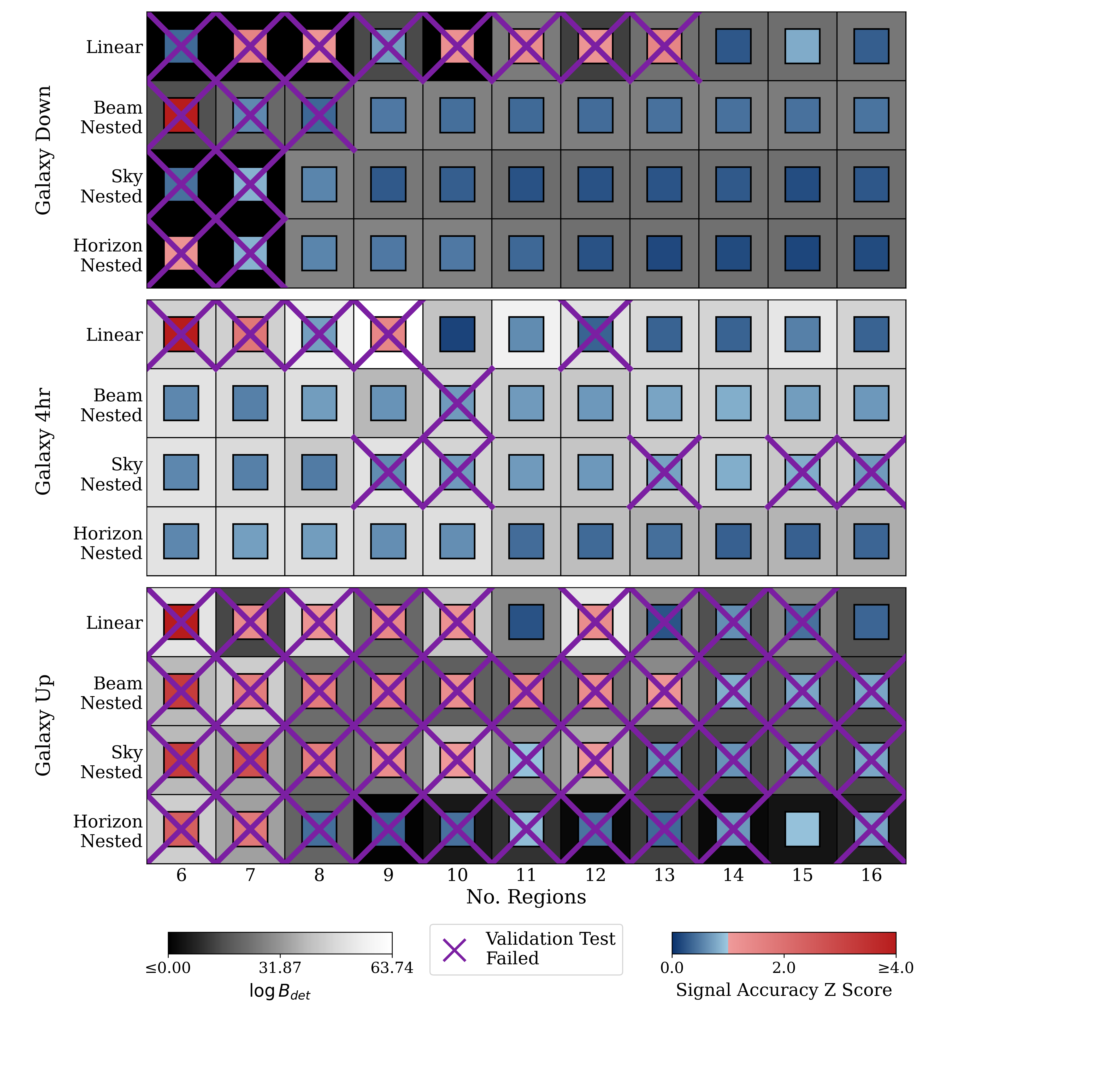}
    \caption[Signal recovery and model selection across sky-splitting methods]{Summary of signal recovery and Bayesian model selection across sky-splitting methods and region counts. Each panel corresponds to a unique observing window, with columns indicating the number of regions ($N_{\rm reg} = 6\text{--}16$) and rows showing the different sky-splitting schemes. The outer shaded square encodes the Bayesian evidence in favour of a signal detection, quantified by the log Bayes factor $\log B_{\rm det}$, with lighter colours indicating stronger support for the signal model. The inner square displays the statistical consistency of the recovered posterior with the injected signal via the Z-score, shown using a blue-to-red colormap. Validation failures are indicated by a single blue cross, drawn when either the null test (signal preferred in no-signal injections) or the residual consistency test fails, signalling potential residual systematics degenerate with the global 21-cm signal.}
    \label{fig:mastergrid}
\end{figure*}

\subsection{Signal Recovery}
\label{sec:signalresults}

Finally, we address the primary objective of this work, the capacity of the proposed modelling frameworks to pass the validation metrics and accurately recover the underlying 21-cm signal. Figure~\ref{fig:mastergrid} summarises the results of the complete suite of nested sampling runs, with each cell reporting the detection Bayes factor ($\ln B_{\mathrm{det}}$), the null-test evidence ratio ($\ln B_{\mathrm{val}}$), and the overall signal recovery accuracy ($Z$-score). Each entry thus encompasses the outcome of four independent inference runs.

These results demonstrate the expected behaviour of the validation framework in a composite model-comparison setting. In particular, they illustrate that a high Bayes ratio favouring a detection ($\ln B_{\mathrm{det}}$) alone is not sufficient to guarantee unbiased signal recovery, and that explicit null-test validation is required to identify models whose component structure permits signal-foreground degeneracies. Across all investigated cases, we observe the framework successfully identifies and flags all cases in which signal recovery is inaccurate while previously $\ln B_{\mathrm{det}}$ in isolation would have suggested decisive statistical support for a detection. Without this additional validation step, one could accept a biased parameter estimate, producing a category II model-comparison problem: a regime in which biased composite models remain degenerate in evidence space with unbiased alternatives (see \citet{Sims_2025} for details).

In analysing signal recovery across the three observational windows, the advantage of sky-aware partitioning is clear. For the Galaxy Down case, the schemes introduced in this work achieve accurate, validated recovery ($Z < 1$) with only 8 regions (sky and horizon-nested) or 9 regions (beam-nested), whereas the observation-independent linear split does not reach validated recovery until 14 regions. The same pattern appears in the 4-hour Galaxy integration, where all sky-aware schemes reach their first sufficient validated configuration at 6 regions (exemplified by the recovered posterior of the beam-nested $N_{\rm reg}=6$ case in Appendix~\ref{app:signalcorner}, Figure~\ref{fig:signal_corner_appendix}), compared with 10 regions for the linear scheme. These shifts demonstrate that the dynamic, importance-aware constructions deliver robust recovery at substantially lower model dimensionality.

We note there exists a small subset of the 4-hour Galaxy integrations at higher $N_{\rm reg}$ for which the recovered signal remains accurate ($Z<1$), yet the configuration is flagged by the null test. This behaviour reflects the existence of signal-foreground degeneracies within the composite model, consistent with the category II scenario. In such cases, the foreground model is sufficiently expressive to recover the modest-amplitude injected fiducial signal when present; however, in signal-free validation data the same flexibility admits regions of parameter space in which the additional signal component yields a marginal improvement in fit, thereby reducing the expected Occam penalty.

Specifically, while there exist regions of parameter space in which the maximum likelihood foreground model yields an adequate description, there also appear to be neighbouring regions of the posterior in which small shifts in spectral-index parameters leave coherent residual structure that can be partially absorbed by the addition of a signal component. This behaviour is consistent with residual signal-foreground degeneracy, in which the effective Occam penalty associated with introducing the signal parameters is reduced relative to the idealised case. In such regimes, the null-test evidence ratio becomes sensitive to the detailed posterior volume structure and to the chromatic interplay between foreground and signal parameterisations. Empirically, we find that these cases correspond to evidence ratios that lie extremely close to the null-test threshold. In practice, the classification can therefore be sensitive to the numerical uncertainty of the nested sampling evidence estimates (see Pattison et al., in prep., for further investigation).

Although this behaviour may give the impression that the validation framework is stringent, such conservatism is essential in composite signal-nuisance inference. In large part, it is a consequence of the prior distributions defined in Table~\ref{tab:priors}, which permit the signal model to explore small amplitudes across a wide range of allowable spectral structure. This strictness ensures that the null-test criterion remains agnostic to the specific shape and structure of the true astrophysical signal. That is, any foreground model that passes the null test and residual checks should permit signal recovery without bias within the stated uncertainties, irrespective of its magnitude (above the noise limit) and its central frequency and characteristic width. The null test therefore operates as a safeguard against residual category II degeneracies, rather than as a performance metric tied to a specific injected signal realisation. Overall, this enables the identification and rejection of high-evidence models that yield biased 21-cm signal recovery at the expense of occasional rejection of otherwise acceptable models.

Finally, we turn to the Galaxy Up case, included throughout this work as a stringent stress-test of the validation framework and of foreground recovery, rather than as an observing window that would be employed for cosmological signal recovery. Physically, this configuration produces the strongest beam--sky coupling and hence the most pronounced chromatic distortions. In direct analogy with the behaviour described above, we find that while some high-$N_{\rm reg}$ configurations yield accurate posterior signal estimates ($Z<1$), they can still be flagged by the null test alone ($\ln B_{\mathrm{val}} \geq 0$). In this maximally chromatic window, the signal--foreground degeneracy is amplified, such that the signal model is more commonly marginally preferred in no-signal injected data even when the accuracy criterion passes and the recovered signal is deemed consistent with the injected realisation. We emphasise that, while the Galaxy Up observing window (restricted here to $N_{\rm reg}\leq 16$) does allow cases in which our fiducial moderate-amplitude injection (150~mK) can be recovered, an astrophysical signal an order of magnitude smaller would not support comparably robust recovery in this observing window. The null-test flagging in these cases is therefore an expected consequence of the amplified chromatic complexity.

As illustrated by Figure~\ref{fig:mastergrid}, multiple configurations may pass the validation criteria for a given observing window. When applied to real observational data, the validation framework therefore need not select a unique foreground configuration. In this case, the final 21-cm constraints should account for the fact that several foreground models remain statistically acceptable, rather than being based on an arbitrary choice of a single validated configuration. A natural route would be Bayesian model averaging, in which the signal posterior is treated as a mixture of the posteriors from all validated configurations with weights set by their evidences.

Motivated by these null-test-driven failures in the maximally chromatic Galaxy Up stress test, an intriguing avenue for future work is to use the null-test Bayes factor itself as an optimisation criterion for region definition, i.e. to directly penalise partitions that admit signal-like fits in no-signal injections. Such an approach would require a parameterised, but not explicitly information-aware, CDF construction. Since exploring $\mathcal{O}(10^2)$ alternative CDF configurations via full nested sampling would be computationally prohibitive even with the GPU-accelerated pipeline presented here, this direction would likely require simulation-based surrogates capable of rapidly forecasting Bayesian model comparison. Recent developments in evidence networks \citep{Jeffrey_2024, Gessey_2024} and conditional Bayesian Neural Ratio Estimation (cBNRE; Leeney et al., in prep) offer a promising route toward enabling such evidence-driven optimisation; a detailed investigation is left to future work.

\section{Conclusions}
\label{sec:conclusions}

In this work, we have presented a significant advancement in the computational and methodological framework for physically-motivated global 21-cm signal analysis. By leveraging GPU architectures and algorithm parallelisation, we achieved a substantial acceleration of Nested Sampling Inference, reducing computational wall-time by factors of $\mathcal{O}(10^{1}$--$10^{3})$. Under the algorithmic configurations used in this work, this enabled the full suite of 1052-run to be completed in under 2 GPU days, compared with an estimated $\sim 100$ CPU core-years, corresponding to an estimated financial saving of over 2 orders of magnitude. This computational efficiency enabled the development and rigorous validation of a novel, observation-dependent sky-partitioning scheme designed to address the challenges of chromatic beam distortions and Galactic foreground contamination. Our results demonstrate that this dynamic partitioning improves foreground modelling through three primary avenues:

\begin{itemize} 
\item \textbf{Principled Model Selection:} The enforcement of a strictly nested region hierarchy allows for the clear identification of the saturation of the maximum log-likelihood, $\ln \mathcal{L}_{\mathrm{max}}$ and consequently the Occam penalty within the Bayesian evidence, $\ln \mathcal{Z}$. This facilitates a statistically robust optimisation of model complexity, ensuring that the number of foreground regions is sufficient to accurately reconstruct the sky to the required precision without unnecessary over-parameterisation.

\item \textbf{Improved Foreground Reconstruction:} The scheme yields more accurate recovery of spatially varying spectral indices. The resulting posterior distributions are consistently centred within true physical ranges, even in challenging observing windows such as when there is maximal coupling between the chromatic beam and high-intensity emission from the Galactic plane.

\item \textbf{Efficient Signal Recovery:} Complex Galactic foregrounds can now be modelled at the precision required for robust global 21-cm signal recovery using a significantly smaller parameter set. This improvement is exemplified in the 4-hour integrated dataset at $N_{\rm reg}=6$: the traditional observation-independent linear sky-parameterisation scheme yields a signal-recovery Z-score of $Z=5.61$, indicative of a highly biased recovery, whereas the beam-nested scheme yields $Z=0.58$, representing an almost order-of-magnitude reduction and bringing the recovery below the $Z<1$ validation threshold. Overall, validated recovery is achieved at lower model complexity in key cases (Galaxy Down: $N_{\rm reg}=8/9$ versus $14$ for linear; 4-hour integration: $6$ versus $10$), corresponding to an approximate 40\% reduction in required foreground-region dimensionality. This reduction, combined with our GPU-accelerated pipeline, makes high-fidelity Bayesian inference far less computationally expensive for large-scale experimental datasets.
\end{itemize}

While this study focused on spectral index masks within the REACH framework, the underlying algorithm is modular and adaptable. It can be readily applied to amplitude-based scale factor maps (e.g. \citet{Pagano_2023}). and is flexible enough to accommodate the diverse beam patterns and instrument responses of different global 21-cm experiments. Future work will explore the integration of this differentiable pipeline into advanced Bayesian and machine learning frameworks, providing a scalable path toward a confirmed detection of the 21-cm signal from the Cosmic Dawn and the Epoch of Reionisation.

\section*{Acknowledgements}

The authors thank Will Handley for his contributions to the REACH pipeline over the years as well as David Yallup for the development of BlackJax's Nested Sampling framework. The authors also thank the reviewer for their valuable feedback on the manuscript. JT is supported by the Harding Distinguished Postgraduate Scholars Programme (HDPSP) and the Science and Technology Facilities Council (STFC) DTP Studentship. We would also like to thank the Kavli Foundation for their support of REACH. EdLA acknowledges the support of STFC via an Ernest Rutherford Fellowship.

\section*{Data Availability}

The data that supports the findings of this study are available from the first author upon reasonable request.



\bibliographystyle{mnras}
\bibliography{references} 




\appendix
\section{Representative Signal-Recovery Posteriors}
\label{app:signalcorner}

\begin{figure*}
    \centering
    \includegraphics[width=\textwidth, trim=0 0 0 0, clip]{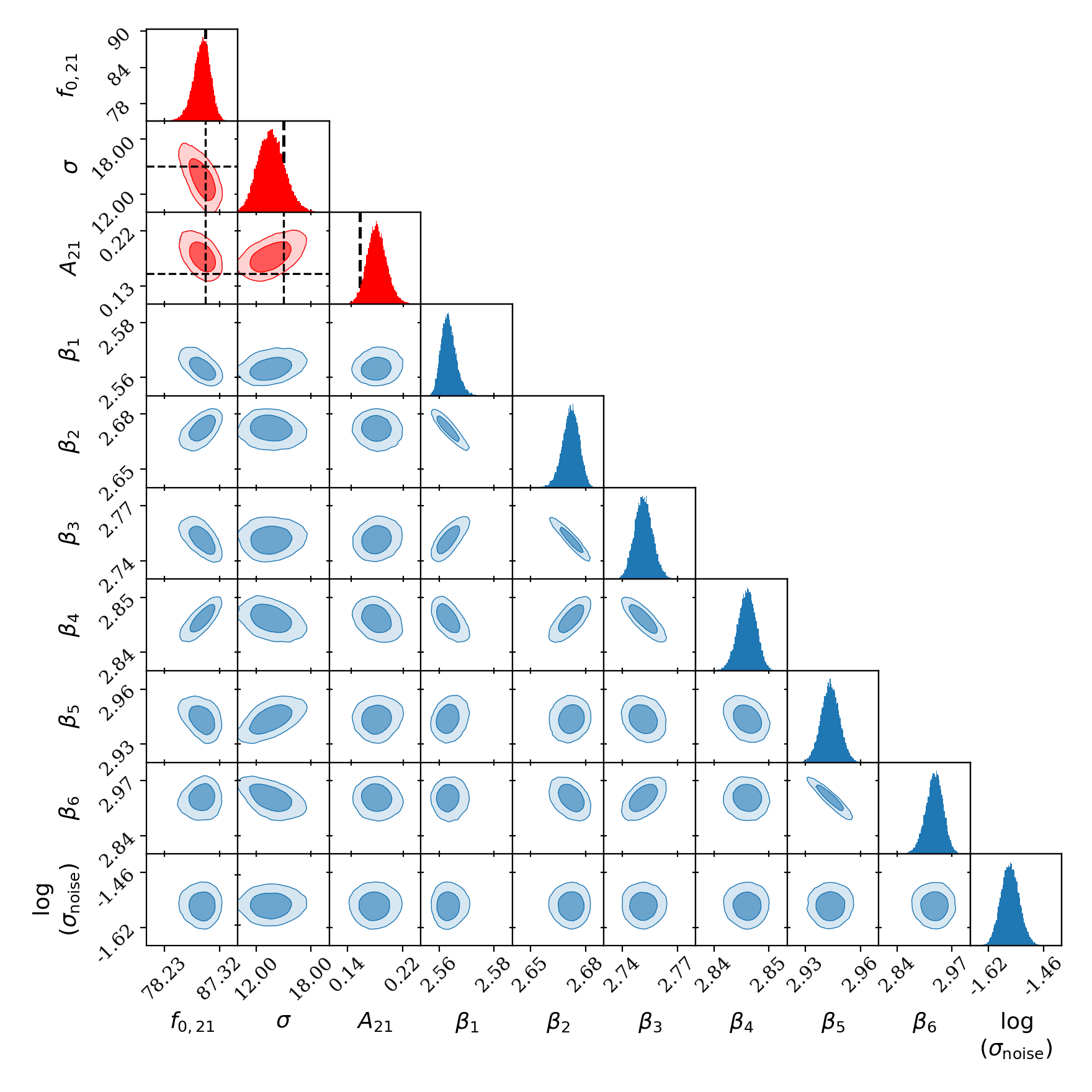}
    \caption{Corner plot of the posterior distributions for the SignalData/SignalFit case of the 4-hour Galactic integration using beam-nested sky partitioning with \(N_{\rm reg}=6\). Signal parameters are shown in red, including the 21-cm amplitude \(A_{21}\), central frequency \(f_0\), and width \(\sigma\), while foreground and noise parameters are shown in blue, including the spectral indices \(\beta_1,\ldots,\beta_6\) and the noise amplitude, \(\log(\sigma_{\mathrm{noise}})\). Diagonal panels show one-dimensional marginalised posteriors, and lower-triangle panels show the two-dimensional posterior structure estimated via kernel density estimation. Black dashed lines mark the injected fiducial signal parameter values \((A_{21}, f_0, \mathrm{std})\) for reference. Generated using the \texttt{anesthetic} package \citep{anesthetic}.}
    \label{fig:signal_corner_appendix}
\end{figure*}

\bsp	
\label{lastpage}
\end{document}